\documentclass[journal,draftclsnofoot,onecolumn,12pt]{IEEEtran}
\usepackage{amsthm,amssymb,graphicx,multirow,amsmath,color,amsfonts,subcaption}
\usepackage[update,prepend]{epstopdf}
\usepackage[noadjust]{cite}
\usepackage[latin1]{inputenc}
\usepackage{tikz}
\usetikzlibrary{arrows,calc}		
\usepackage{bbm} 
\usepackage{pdfpages}
\usepackage{tabulary}
\usepackage{multirow}
\usepackage{comment}

\def\nba{{\mathbf{a}}}

\def\nbo{{\mathbf{o}}}
\def\nbp{{\mathbf{p}}}

\def\nbu{{\mathbf{u}}}

\def\nb0{{\mathbf{0}}}
\def\nb1{{\mathbf{1}}}

\def\nsfA{{\mathsf{A}}}
\def\nsfB{{\mathsf{B}}}

\def\nsfG{{\mathsf{G}}}
\def\nsfH{{\mathsf{H}}}

\def\nsfR{{\mathsf{R}}}
\def\nsfS{{\mathsf{S}}}
\def\nsfT{{\mathsf{T}}}

\def\nsfV{{\mathsf{V}}}
\def\nsfW{{\mathsf{W}}}
\def\nsfX{{\mathsf{X}}}
\def\nsfY{{\mathsf{Y}}}

\def\nsfTheta{\mathsf{\Theta}}

\def\ncalA{{\mathcal{A}}}

\def\nbbC{{\mathbb{C}}}

\def\nbbE{{\mathbb{E}}}

\def\nbbP{{\mathbb{P}}}

\def\nbbR{{\mathbb{R}}}

\newtheorem{lemma}{Lemma}

\newtheorem{theorem}{Theorem}

\newtheorem{cor}{Corollary}

\newtheorem{remark}{Remark}

\newtheorem{approximation}{Approximation}
	
\def\SPEB{\nsfS(\underline{\nsfR}^{(N)},\underline{\nsfTheta}^{(N)})}
\def\GDOP{\nsfG(\nsfR,\underline{\nsfTheta}^{(N)})}
\allowdisplaybreaks 
\begin{document}
\title{Characterizing the Impact of SNR Heterogeneity on Time-of-Arrival based Localization Outage Probability}
\author{Sundar Aditya,~\IEEEmembership{Student~Member,~IEEE}, Harpreet S. Dhillon,~\IEEEmembership{Member,~IEEE}, Andreas F. Molisch,~\IEEEmembership{Fellow,~IEEE}, R. Michael Buehrer,~\IEEEmembership{Fellow,~IEEE} and Hatim Behairy
\thanks{Sundar Aditya and Andreas F. Molisch are with WiDeS, Ming Hsieh Dept. of Electrical Engineering, University of Southern California, Los Angeles, CA 90089, USA (Email:\{sundarad, molisch\}@usc.edu).}
\thanks{Harpreet S. Dhillon and R. Michael Buehrer are with Wireless@VT, Bradley Dept. of Electrical and Computer Engineering, Virginia Tech., Blacksburg, VA 24061, USA (Email: \{hdhillon, buehrer\}@vt.edu).}
\thanks{Hatim Behairy is with King Abdulaziz City for Science and Technology, P. O. Box 6086, Riyadh 11442, Saudi Arabia (Email: hbehairy@kacst.edu.sa) \hfill Revised: \today.}
\thanks{This work was supported by KACST under grant number 33-878.}
}

\maketitle

\begin{abstract}
In localization, an outage occurs if the positioning error exceeds a pre-defined threshold, $\epsilon_{\rm th}$. For time-of-arrival based localization, a key factor affecting the positioning error is the relative positions of the anchors, with respect to the target location. Specifically, the positioning error is a function of (a) the \emph{distance-dependent} signal-to-noise ratios (SNRs) of the anchor-target links, and (b) the pairwise angles subtended by the anchors at the target location. From a design perspective, characterizing the distribution of the positioning error over an ensemble of target and anchor locations is essential for providing probabilistic performance guarantees against outage. To solve this difficult problem, previous works have assumed all links to have the same SNR (i.e., SNR homogeneity), which neglects the impact of distance variation among the anchors on the positioning error. In this paper, we model SNR heterogeneity among anchors using a distance-dependent pathloss model and derive an accurate approximation for the error complementary cumulative distribution function (ccdf). By highlighting the accuracy of our results, relative to previous ones that ignore SNR heterogeneity, we concretely demonstrate that SNR heterogeneity has a considerable impact on the error distribution.
\end{abstract}


\section{Introduction} \label{sec:intro}
\begin{figure}
  \centering
   \includegraphics[scale=0.35]{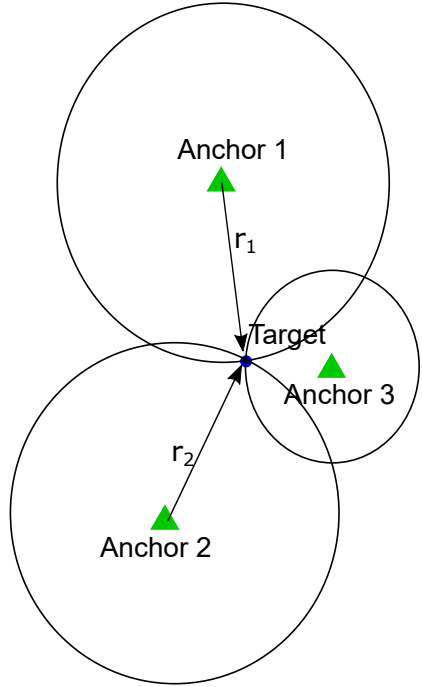}
   \caption{ToA based localization: Each distance (range) estimate constrains the target to lie on a circle centered at the corresponding anchor, whose radius equals the range. The intersection of three or more such circles provides an unambiguous solution for the target location in $\nbbR^2$.}
   \label{fig:trilateration}
\end{figure}

In recent years, the number of applications requiring accurate position information has grown steadily; from navigation for autonomous vehicles \cite{Cui_Xue_Zheng_2016} to crowd-sensing \cite{Cardone_et_al_2013}, location-based advertising \cite{Steiniger_et_al_2006} and virtual reality \cite{Clark_2016}, to name a few. On a two-dimensional surface, a target can be localized if its distance (also known as \emph{range}) to at least three fixed reference points, called anchors, is known (see Fig.~\ref{fig:trilateration}). For wireless systems, ranges can be estimated from the time-of-arrival (ToA) of a \emph{ranging signal}\footnote{This requires the targets and anchors to be synchronized.} and for wideband systems in particular, ToA-based localization is especially attractive, since the finer time resolution due to the large bandwidth improves the accuracy of the range estimates \cite{Gezici_et_al_2005}, thereby resulting in accurate location estimates.

While the principle behind ToA-based localization is fairly straightforward, a variety of operating conditions and propagation phenomena, such as noise, interference, multipath, blocking, target mobility etc. render the task of designing a localization network challenging. In order to provide a reliable quality of service in terms of accuracy (e.g., an error of at most $50{\rm m}$ at least $90\%$ of the time, as mandated by the the E911 standard \cite{e911_2015}), it is important to characterize the probability distribution of the positioning error over an ensemble of operating conditions, especially for safety critical applications like autonomous vehicles or E911 emergency services. A commonly used metric for this purpose is the localization mean-squared error (MSE), which is a function of the anchor locations, the transmit powers, the propagation environment, as well as the choice of ranging and localization algorithms. In this work, we consider the impact of the anchor locations, relative to a target, on the MSE. Specifically, we consider a lower bound for the MSE, known as the squared position error bound (SPEB) \cite{Jourdan_Dardari_Win_2008,Shen_Win_2010}, which is satisfied by all positioning algorithms that return unbiased\footnote{An estimate $\hat{\nbp}$ of a target location $\nbp$ is said to be unbiased if $\nbbE[\hat{\nbp}]=\nbp$.} estimates of the target location. The SPEB is a function of the anchor geometry and importantly, does not depend on a specific localization algorithm. As a result, it is well-suited as a metric to analyze the impact of the anchor geometry on the positioning error. If the SPEB exceeds a pre-defined threshold, $\epsilon_{\rm th}$, then the target is said to be in \emph{outage}. Over an ensemble of target and/or anchor locations, the SPEB (and the MSE, as well) is a random variable and characterizing its complementary cumulative distribution function (ccdf) in closed-form (i.e., $\nbbP(\mbox{SPEB} > u)$, as a function of $u$) is important from a design perspective, as it can be used to determine a deployment of anchors that can guarantee an outage probability of at most $p_{\rm out}$\footnote{For a given error threshold, $\epsilon_{\rm th}$, an outage probability of at most $p_{\rm out}$ can be guaranteed if and only if the condition $\nbbP(\mbox{SPEB} > \epsilon_{\rm th}) \leq p_{\rm out}$ is satisfied, which poses a constraint on the shape of the SPEB ccdf.}.

Given $N$ anchors in a region, a natural model for capturing the randomness in the anchor locations is the well-known binomial point process (BPP) \cite[Chap.~2]{Sto_et_al_full_2013}, in which the anchors are distributed independently and uniformly over the region. In this paper, we attempt to derive a closed-form expression for the SPEB ccdf for such an anchor model\footnote{Typically, the number of anchors, $N$, is also a random variable, often modeled as having a Poisson distribution. Together with the randomness in the anchor locations, this corresponds to the well-known homogeneous Poisson point process (PPP), which has been used to analyze the localization performance of a variety of wireless networks \cite{Schlo_Dhill_Buehr_2015, Bhandari_Dhillon_Buehrer_2016, Schlo_Dhill_Buehr_2016_2, Adi_Har_Mol_Beh_2017, Adi_Har_Mol_Beh_2017_2}. Hence, the results presented in this paper for the BPP anchor model can be readily extended for the PPP case by averaging over the distribution of $N$.}. Our approach is summarized below.

\subsection{Methodology} \label{sec:contributions}
\begin{itemize}
 \item For a given target, we assume that the anchors that are within its communication range are distributed according to a BPP over an annular region centered at the target. For this setup, we model the SNR heterogeneity across different anchor-target links using a pathloss model. As a result, the SPEB metric is a function of the anchor distances and angular positions, relative to the target.
 \item Given $N$ anchors, we rearrange the SPEB expression and reduce it in terms of the product of two random variables, $\nsfX_N$ and $\nsfY_N$. While $\nsfX_N$ depends only on the anchor distances, $\nsfY_N$ depends on both the distances and angular positions of the anchors. In particular, $\nsfY_N$ and $\nsfX_N$ are statistically dependent.
 \item We then proceed to demonstrate that the conditional distribution of $\nsfY_N$, given $\nsfX_N$, is difficult to characterize in closed-form. Hence, through \emph{constrained moment matching}, we derive an approximation for $\nsfY_N$, denoted by $\nsfV_N$, which depends only on the angular positions of the anchors and has the same mean as $\nsfY_N$. In particular, $\nsfX_N$ and $\nsfV_N$ are statistically independent.
 \item Consequently, the SPEB can be approximated in terms of the product of independent random variables, $\nsfX_N$ and $\nsfV_N$, and we derive a closed-form expression for the ccdf of this approximation (see Theorem~\ref{thm:speb_dist} in Section~\ref{sec:speb_distb.}), which is the key result of this paper.
\item Through simulations, we verify that the derived SPEB ccdf accurately estimates the true ccdf. Thus, from a design perspective, our contribution is useful in determining the number of anchors required in order to satisfy $p_{\rm out}\leq \delta$, for any $\delta \in (0,1)$. We also show that the accuracy of our approach is superior to that of other approaches that ignore SNR heterogeneity, which serves to highlight the impact of SNR heterogeneity on the SPEB (and consequently, the MSE) distribution.
\end{itemize}

\subsection{Related Work}  \label{sec:related}
There have been a number of recent works that have focused on the impact of anchor geometry on the localization error performance; specifically, for the SPEB metric, the impact of the target being situated within the convex hull of the anchors was investigated in \cite{Olone_Buehrer_2016}, while scaling laws, with respect to the number of anchors within communication range, were derived in \cite{Shen_et_al_2010}. A related, but simpler metric, known as the geometric dilution of precision (GDOP) has been studied extensively for the BPP anchor model. The GDOP corresponds to a special case of the SPEB when all the anchor-target links have the same SNR. The asymptotic distribution of the GDOP, as the number of anchors approaches infinity, was derived using U-statistics in \cite{Huang_etal_2012}. For the more realistic case of a finite number of anchors, the max-angle metric was proposed and analyzed in \cite{Schlo_Dhill_Buehr_2016} and shown to be correlated to the GDOP. An approximate GDOP distribution was presented in \cite{Olone_Dhillon_Buehrer_2017}, using the order statistics of the inter-node angles, while the \emph{exact} GDOP distribution was characterized in \cite{Zhou_Shen_2017}. To the best of our knowledge, ours is the first work to consider the more realistic scenario where the anchor-target links may have different SNRs (due to the anchors being situated at different distances from the target), which increases the difficulty of the problem considerably, as highlighted in Section~\ref{sec:SysMod}. 

\subsection{Notation}
Throughout this work, bold lower case letters are used for deterministic vectors. In particular, $\nb1$ denotes the all-one vector. Uppercase letters in serif font are used for scalar random variables (e.g., $\mathsf{X}$), while random vectors are underlined and similarly represented (e.g., $\underline{\mathsf{X}}$). For square matrices, the trace and inverse operators are respectively denoted by ${\rm tr(\cdot)}$ and $(\cdot)^{-1}$. $\nbbR$ represents the real numbers, $\nbbC$ the complex numbers, $i\in \nbbC$ the imaginary unit, and ${\rm Im}(z)$ the imaginary part of $z \in \nbbC$. For random variables $\nsfX$ and $\nsfY$, $f_\nsfX(\cdot)$, $\overline{F}_\nsfX(\cdot)$ and $\varphi_\nsfX(\cdot)$ denote the marginal probability density function (pdf), the marginal ccdf and the characteristic function of $\nsfX$, respectively, while $\overline{F}_{\nsfX|\nsfY}(.|y)$ denotes the conditional ccdf of $\nsfX$, given $\nsfY=y$. $\nbbP(.)$ denotes the probability measure, while $\nbbE_{\nsfX}[.]$ denotes the expectation operator over the distribution of $\nsfX$. A real, parametrized function $h:\nbbR \rightarrow \nbbR$, with argument $t$ and parameters given by a vector, $\nba$, is denoted by $h(t;\nba)$. For $x\in \nbbR$, the sine and cosine integrals, denoted by ${\rm Si}(x)$ and ${\rm Ci}(x)$, respectively, are defined as follows:
\begin{align}
\label{eq:Si}
  {\rm Si}(x) &= \displaystyle\int\limits_0^x \frac{\sin t}{t}~ {\rm d}t, \\
\label{eq:Ci}  
  {\rm Ci}(x) &= - \displaystyle\int\limits_x^\infty \frac{\cos t}{t}~ {\rm d}t.
\end{align} 
$\mathbbm{1}(\cdot.)$ denotes the indicator function and finally, the function ${\rm H}:\nbbR \rightarrow \nbbC$ is defined as follows:
\begin{align}
\label{eq:H}
   {\rm H}(x) &:= {\rm Si}(x)-i{\rm Ci}(x), ~ x\in \nbbR.
\end{align}

\subsection{Organization}
This paper is divided into five sections. The system model is described in Section~\ref{sec:SysMod}, where the anchors are modeled by a BPP over an annular region surrounding the target, and a distance-dependent pathloss model is assumed for the SNRs of the anchor-target links. Under these conditions, we illustrate the difficulty of characterizing the SPEB distribution in Section~\ref{sec:speb_distb.}, which motivates the derivation of a tractable approximation for the SPEB ccdf later on in the same section. In Section~\ref{sec:NumResults}, we compare the accuracy of our approach with other bounds and approximations that do not consider SNR heterogeneity. Finally, Section~\ref{sec:concl} concludes the paper.

\section{System Model} \label{sec:SysMod}
Consider a target situated in $\nbbR^2$ that needs to be localized. Since we are interested in the anchor geometry relative to the target, we can assume, without loss of generality, that the target is situated at the origin, $\nbo$. Centered at the target, consider $N\geq 3$ anchors deployed according to a BPP over an annular region from $d_{\rm min}$ to $d_{\max}~ (d_{\rm max}>d_{\rm min}>0)$\footnote{$d_{\rm max}$ can be interpreted as the distance beyond which $s(t)$ is too weak to be detected by the target.}, denoted by $\ncalA_\nbo(d_{\rm min},d_{\rm max})$, and let $(\nsfR_k,\nsfTheta_k)$ denote the location of the $k$-th anchor in polar coordinates $(\nsfR_k \in [d_{\rm min}, d_{\rm max}], ~\nsfTheta_k \in [0,2\pi); ~ k=1, \cdots, N)$. Let $s(t)$, having Fourier transform $S(f)$, denote the ranging signal transmitted by the anchors\footnote{We assume that the anchors coordinate their transmissions to avoid interference at the target. As a result, ToA/range estimation is noise-limited.} and let $y(t;\nsfR_k,\nsfTheta_k)$ denote the signal received from the $k$-th anchor, which can be modeled as a superposition of a number of multipath components (MPCs) in the following manner:  
\begin{align}
 \label{eq:y_ac}
 y(t;\nsfR_k,\nsfTheta_k) &= \displaystyle\sum\limits_{l=1}^{L(\nsfR_k,\nsfTheta_k)} \alpha_{l}(\nsfR_k,\nsfTheta_k) s(t-\tau_{l}(\nsfR_k,\nsfTheta_k)) + \eta_{k}(t),~ k\in\{1,\cdots,N\},
\end{align}
where the location-dependent quantities $L(\nsfR_k,\nsfTheta_k)$, $\alpha_{l}(\nsfR_k,\nsfTheta_k) \in \nbbC$ and $\tau_{l}(\nsfR_k,\nsfTheta_k)\in \nbbR$ respectively denote the number of observed MPCs, the complex amplitude of the $l$-th MPC and its ToA. $\eta_{k}(t)$ is the measurement noise, which is modeled as a zero-mean complex Gaussian random process, having a power spectral density of $N_0$. We assume that line-of-sight exists from the target to all the anchors. Hence, the first arriving MPC from each anchor corresponds to the direct path (DP) and depends on the anchor position as follows: 
\begin{align}
 \label{eq:tau}
 \tau_{1}(\nsfR_k,\nsfTheta_k)= \tau_{1}(\nsfR_k) = \frac{\nsfR_k}{c},~ k\in\{1,\cdots,N\}, 
\end{align}
where $c$ denotes the speed of light in free space. The other MPCs are known as indirect paths (IPs) and we assume no prior knowledge of their statistics. Under these conditions, the MSE of an unbiased estimate of the target location can be bounded using the Cramer-Rao lower bound (CRLB) \cite{Jourdan_Dardari_Win_2008,Shen_Win_2010}, as follows:
\begin{align}
 \label{eq:mse}
 {\rm MSE} &\geq {\rm tr}\left(\left( \displaystyle\sum\limits_{k=1}^{N} \mu(\nsfR_k,\nsfTheta_k) \nbu(\nsfTheta_k) \nbu(\nsfTheta_k)^T\right)^{-1}\right) \\
 \label{eq:speb}
 &:= \SPEB, \\
\label{eq:Rsf}
\mbox{where}~ \underline{\nsfR}^{(N)} &= [\nsfR_1,~ \cdots,~\nsfR_N]^T, \\
\label{eq:thetasf}
\underline{\nsfTheta}^{(N)} &= [\nsfTheta_1,~ \cdots,~\nsfTheta_N]^T, \\ 
\label{eq:mu}
 \mu(\nsfR_k,\nsfTheta_k) &= \frac{8 \pi^2 \beta^2 (1-\chi(\nsfR_k,\nsfTheta_k)) \gamma(\nsfR_k,\nsfTheta_k)} {c^2}, \\
 \label{eq:snr}
 \gamma(\nsfR_k,\nsfTheta_k) &= \frac{|\alpha_{1}(\nsfR_k,\nsfTheta_k)|^2}{N_0}\displaystyle\int\limits_{-\infty}^{\infty} |s(t)|^2~ {\rm d}t,\\
 \label{eq:beta}
 \beta &= \left[\left(\displaystyle\int\limits_{-\infty}^\infty f^2 |S(f)|^2 {\rm d}f\right) \middle/ \left(\displaystyle\int\limits_{-\infty}^\infty |S(f)|^2 {\rm d}f \right)\right]^{1/2}, \\
\mbox{and}~ \nbu(\nsfTheta_k) &= [\cos(\nsfTheta_k) ~ \sin(\nsfTheta_k)]^T.
\end{align}
$\SPEB$ is commonly known as the squared-position error bound (SPEB) \cite{Jourdan_Dardari_Win_2008,Shen_Win_2010} in localization terminology. The term $\mu(\nsfR_k,\nsfTheta_k)$ is referred to as the \emph{ranging information intensity} (RII) from the $k$-th anchor and is a measure of the ranging accuracy associated with the $k$-th anchor\footnote{$\mu(\nsfR_k,\nsfTheta_k)$ is the reciprocal of the CRLB for an unbiased estimate of $\nsfR_k$ \cite{Gezici_et_al_2005}.}. It is a function of the DP SNR, $\gamma(\nsfR_k,\nsfTheta_k)$, the \emph{effective bandwith}, $\beta$, and the path overlap factor, $\chi(\nsfR_k,\nsfTheta_k) \in [0,1]$, which determines the extent of overlap between the DP and subsequent MPCs, due to finite bandwidth\footnote{The expression for $\chi(\nsfR_k,\nsfTheta_k)$ can be found in \cite{Shen_Win_2010}.}. For simplicity, we assume $\chi(\nsfR_k,\nsfTheta_k)=0$ for all $k$, which corresponds to the case where the DP does not overlap with any other MPC, thereby resulting in the most accurate estimate of $\nsfR_k$. Furthermore, $\gamma(\nsfR_k,\nsfTheta_k)$ is a function of the DP attenuation, $|\alpha_{1}(\nsfR_k,\nsfTheta_k)|^2$, for which the following \emph{pathloss} model is assumed: 
\begin{align}
 \label{eq:pathloss}
 |\alpha_{1}(\nsfR_k,\nsfTheta_k)|^2 = |\alpha_{1}(\nsfR_k)|^2 = (d_{\rm min}/\nsfR_k)^2 .
\end{align}
\begin{remark}
For anchors having line-of-sight to the target, the inverse-square law pathloss model in (\ref{eq:pathloss}) is a reasonable assumption for the DP component if there is zero path overlap, which, in turn, can be assumed when $d_{\rm max} < d_{\rm break}$, where $d_{\rm break}$ denotes the breakpoint distance associated with the ground reflection \cite{Molisch_book_2011}, since zero overlap between the DP and the ground-reflected path can be rarely achieved.
\end{remark}

Apart from the RIIs, which depend primarily on the ranges, $\SPEB$ also depends on the angular geometry of the anchors, which is captured in (\ref{eq:mse}) by the outer product $\nbu(\nsfTheta_k)\nbu(\nsfTheta_k)^T$, where $\nbu(\nsfTheta_k)$ is the unit vector in the direction of the $k$-th anchor. In summary, the $k$-th term in the summation in (\ref{eq:mse}) represents the contribution of the $k$-th anchor to $\SPEB$. From (\ref{eq:mse})-(\ref{eq:pathloss}), $\SPEB$ can be expressed as follows:
\begin{align}
\label{eq:speb_exp}
\SPEB &= \frac{\displaystyle\sum\limits_{k=1}^N \nsfR_k^{-2}}{T_{\rm s}\displaystyle\sum\limits_{j=1}^{N-1} \displaystyle\sum\limits_{k=j+1}^{N} \nsfR_j^{-2} \nsfR_k^{-2} \sin^2(\nsfTheta_j - \nsfTheta_k)}, \\
\mbox{where}~ T_{\rm s} &= \frac{8 \pi^2 \beta^2 d_{\rm min}^{2}}{N_0c^2} \displaystyle\int\limits_{-\infty}^{\infty} |s(t)|^2~{\rm d}t.
\end{align}
Since $\SPEB$ does not depend on any particular positioning algorithm, it is well-suited as a metric to analyze the impact of anchor geometry on the MSE. Moreover, many positioning algorithms have been proposed in recent years that have been shown to satisfy (\ref{eq:mse}) with equality \cite{Shen_et_al_2012, Meissner_Leitinger_Witrisal_2014, Leitinger_Froehle_et_al_2014, Meissner_et_al_2014}. Hence, for the remainder of this paper, we assume that the MSE is identical to $\SPEB$.

For the special case when all the anchors are at the same distance $\nsfR$ from the target (i.e., all links have the same SNR), $\SPEB$ reduces to another well-known metric called the Geometric Dilution of Precision (GDOP), which is denoted by $\GDOP$ and has the following expression\footnote{Technically, the GDOP is defined as the square root of $\GDOP$ \cite{Torrieri_1984} and thus, has the units of distance. However, in order to have a fair comparison with $\SPEB$ (which has units of distance-squared), we slightly abuse the notation and refer to $\GDOP$ as the GDOP in this work.}:
\begin{align}
 \label{eq:gdop}
 \GDOP &= \frac{N\nsfR^{2}}{T_{\rm s} \displaystyle\sum\limits_{j=1}^{N-1} \displaystyle\sum\limits_{k=j+1}^{N} \sin^2 (\nsfTheta_j - \nsfTheta_k)}=\frac{1}{T_{\rm s}}\nsfG_1(\nsfR)\nsfG_2(\underline{\nsfTheta}^{(N)}),\\
 \mbox{where}~ \nsfG_1(\nsfR) &= \nsfR^2, \\
 \mbox{and}~ \nsfG_2(\underline{\nsfTheta}^{(N)}) &=  \frac{N}{\displaystyle\sum\limits_{j=1}^{N-1} \displaystyle\sum\limits_{k=j+1}^{N} \sin^2 (\nsfTheta_j - \nsfTheta_k)}.
\end{align}
Compared to $\SPEB$, $\GDOP$ is more tractable for a statistical characterization, since it can be decomposed into a product of two independent random variables, $\nsfG_1(\nsfR)$ and $\nsfG_2(\underline{\nsfTheta}^{(N)})$, as shown in (\ref{eq:gdop}). However, since the $\sin^2(\cdot)$ terms are weighted differently in the denominator of (\ref{eq:speb_exp}), it is, in general, not possible to express $\SPEB$ as $\nsfS_1(\underline{\nsfR}^{(N)})\nsfS_2(\underline{\nsfTheta}^{(N)})$, for some $\nsfS_1(\cdot)$ and $\nsfS_2(\cdot)$, in much the same way as it is generally not possible to represent an expression like $a_1 x_1+\cdots+a_Mx_M$ as $h_1(a_1,\cdots,a_M)h_2(x_1,\cdots,x_M)$, for some scalar-valued real functions, $h_1(\cdot)$ and $h_2(\cdot)$ and arbitrary real values of $a_i$ and $x_i ~(i=1,\cdots,M)$. Hence, for the sake of tractability, we formulate an approximation
that allows a decomposition of $\SPEB$, along the lines of (\ref{eq:gdop}), in the following section.

\section{Characterizing SPEB distribution}
 \label{sec:speb_distb.}
Although $\SPEB$ cannot, in general, be decomposed as a product of independent random variables, a partial decomposition can be obtained as shown in the lemma below:
\begin{lemma}
\label{lem:speb_new}
The expression for $\SPEB$ in (\ref{eq:speb_exp}) can re-written as follows:
\begin{align}
 \label{eq:speb_alt}
 \SPEB &= \frac{4}{T_{\rm s} \nsfX_N \nsfY_N},\\
 \label{eq:XN}
 \mbox{where}~ \nsfX_N &= \displaystyle\sum\limits_{k=1}^N \nsfA_k, \\
 \label{eq:Ak}
 \nsfA_k &= \nsfR_k^{-2}, \\ 
 \label{eq:YN}
 \nsfY_N &= 1- \left(\displaystyle\sum\limits_{k=1}^N \nsfB_{k,N} \cos 2\nsfTheta_k \right)^2 - \left(\displaystyle\sum\limits_{k=1}^N \nsfB_{k,N} \sin 2\nsfTheta_k \right)^2, \\
 \label{eq:Bkn}
\mbox{and}~ \nsfB_{k,N} &= \frac{\nsfA_k}{\nsfX_N}, ~k=1,\cdots,N.
\end{align}
\end{lemma}
\begin{IEEEproof}
See Appendix \ref{app:speb_new}.
\end{IEEEproof}
While $\nsfX_N$ depends only on $\underline{\nsfR}^{(N)}$, $\nsfY_N$ is a function of both $\underline{\nsfR}^{(N)}$ and $\underline{\nsfTheta}^{(N)}$. Moreover, since $\nsfY_N$ is a function of $\nsfX_N$, the two random variables are statistically dependent. Let $\overline{F}_{\rm S}(\cdot)$ denote the ccdf of $\SPEB$, which can be expressed as follows:
\begin{align}
 \label{eq:speb_tail}
 \overline{F}_{\rm S}(u) &= \nbbP(\SPEB > u) \notag\\
 &= \nbbP\left(\nsfX_N \nsfY_N \leq \frac{4}{uT_{\rm s}} \right) \notag \\
 &= 1-\nbbE_{\nsfX_N}\left[\overline{F}_{\nsfY_N|\nsfX_N}\left(\frac{4}{uxT_{\rm s}}\bigg| x\right)\right]. 
\end{align}

Before proceeding to derive an expression for $\overline{F}_{\rm S}(\cdot)$, we consider the GDOP special case, for which the evaluation of (\ref{eq:speb_tail}) is relatively simpler.
\begin{cor}
For the special case when $\underline{\nsfR}^{(N)}=\nsfR\nb1$ in (\ref{eq:speb_alt})-(\ref{eq:Bkn}), $\SPEB$ reduces to $\GDOP$, which can be re-written as follows:
\begin{align}
 \nsfS(\nsfR\mathbf{1},\underline{\nsfTheta}^{(N)}) &= \GDOP \notag\\
  \label{eq:gdop_alt}
 &= \frac{4 \nsfR^2}{T_{\rm s} N \nsfW_N}, \\
 \label{eq:WN}
 \mbox{where}~ \nsfW_N &= 1- \frac{1}{N^2}\left[\left(\displaystyle\sum\limits_{k=1}^N \cos 2\nsfTheta_k \right)^2 + \left(\displaystyle\sum\limits_{k=1}^N  \sin 2\nsfTheta_k \right)^2 \right].
\end{align}
\end{cor}
From (\ref{eq:gdop_alt}), the ccdf of $\GDOP$, denoted by $\overline{F}_{\rm G}(\cdot)$, can be obtained as follows:
\begin{align}
 \label{eq:gdop_tail}
 \overline{F}_{\rm G}(u) &= \nbbP(\GDOP > u) \notag\\
 &= 1-\nbbE_{\nsfR}\left[\overline{F}_{\nsfW_N|\nsfR}\left(\frac{4r^2}{T_{\rm s}Nu} \bigg| r\right)\right]\notag \\
 &=1-\nbbE_{\nsfR}\left[\overline{F}_{\nsfW_N}\left(\frac{4r^2}{T_{\rm s}Nu}\right)\right], 
\end{align}
where (\ref{eq:gdop_tail}) follows from the independence of $\nsfR$ and $\nsfW_N$. As a result, the marginal distributions of $\nsfR$ and $\nsfW_N$ completely characterize $\overline{F}_{\rm G}(\cdot)$. In particular, the ccdf of $\nsfW_N$ has the following expression \cite{Zhou_Shen_2017},
\begin{align}
 \label{eq:Bessel}
 \overline{F}_{\nsfW_N}(u) &= \begin{cases}
 1,~ u< 0\\
 N\sqrt{1-u} \displaystyle\int\limits_0^\infty J_1\left(N\sqrt{1-u}~ y \right) ~ (J_0(y))^N {\rm d}y,~ u\in[0,1],\\
 0,~ u>1 
 \end{cases}
 \end{align}
where $J_0(.)$ and $J_1(.)$ denote the zeroth and first order Bessel functions, respectively, while the pdf of $\nsfR$ is given by
\begin{align}
 \label{eq:fr}
f_{\nsfR}(r) &= \frac{2r}{d_{\rm max}^2 - d_{\rm min}^2} \mathbbm{1}(r \in [d_{\rm min},d_{\rm max}]).
\end{align}

\begin{remark}
\label{rem:Chris}
Given $N$, the support (i.e., the feasible set of values) of $\SPEB$ and $\GDOP$ is $[4d_{\rm min}^2/(NT_{\rm s}), \infty)$, where the minimum value, $4d_{\rm min}^2/(NT_{\rm s})$, is attained when the anchors are located at the vertices of a regular $N$-sided polygon inscribed within a circle of radius $d_{\rm min}$. Due to the common support, $\overline{F}_{\rm G}(\cdot)$ can be interpreted as a GDOP-based approximation of $\overline{F}_{\rm S}(\cdot)$, where the averaging over $\nsfR$ in (\ref{eq:gdop_tail}) partially takes into account the SNR heterogeneity, while retaining the GDOP structure. In \cite{Olone_Dhillon_Buehrer_2017}, the authors considered an alternate GDOP-based approximation for $\SPEB$, given below, using the average link SNR:
\begin{align}
  \SPEB &\approx  \nsfH(\underline{\nsfTheta}^{(N)}) \notag \\
  &:= \frac{4}{T_{\rm s}N \nbbE[\nsfR^{-2}]\nsfW_N} \notag \\
  \label{eq:Chris}
  &= \nsfG((\nbbE[\nsfR^{-2}])^{-1/2},\nsfTheta^{(N)}),
 \end{align}
where $\nbbE[\nsfR^{-2}]$ is proportional to the average link SNR due to the pathloss model assumed in (\ref{eq:pathloss}). While $\nsfH(\underline{\nsfTheta}^{(N)})$ also partially accounts for SNR heterogeneity, its minimum value is $4/(NT_{\rm s}\nbbE[\nsfR^{-2}]) \geq 4d_{\rm min}^2/(NT_{\rm s})$, since $\nsfR^{-2} \leq d_{\rm min}^{-2}$. To illustrate the impact of this support mismatch, consider the ratio between the minimum values of $\nsfH(\underline{\nsfTheta}^{(N)})$ and $\SPEB$, provided below:
\begin{align}
\label{eq:support_mismatch}
\frac{\displaystyle\min\limits_{\nsfTheta^{(N)}}~ \nsfH(\underline{\nsfTheta}^{(N)})}{\displaystyle\min\limits_{\nsfR^{(N)},\nsfTheta^{(N)}} \SPEB} = \frac{1}{d_{\rm min}^2 \nbbE[\nsfR^{-2}]} = \frac{(d_{\rm max}/d_{\rm min})^2-1}{2 \log(d_{\rm max}/d_{\rm min})}. 
\end{align}
Since (\ref{eq:support_mismatch}) is increasing in $(d_{\rm max}/d_{\rm min})$, we can reasonably conclude that the approximation given by (\ref{eq:Chris}) is unlikely to be accurate when the difference between $d_{\rm min}$ and $d_{\rm max}$ is large. Note that (\ref{eq:support_mismatch}) holds for the inverse-square law pathloss model only.
\end{remark}

We now focus our attention back to the general case of deriving a closed-form expression for $\overline{F}_{\rm S}(\cdot)$ from (\ref{eq:speb_tail}), by characterizing the marginal distribution of $\nsfX_N$ and the conditional distribution $\nsfY_N$, given $\nsfX_N$.

\begin{lemma}
\label{lem:charAk}
The characteristic function of $\nsfX_N$ is given by:
\begin{align}
\label{eq:charXN2} 
\varphi_{\nsfX_N}(t) &= (\varphi_{\nsfA_1}(t))^N, \\
\label{eq:charA2}
\mbox{where}~ \varphi_{\nsfA_1}(t)  &= \frac{1}{d_{\rm max}^2 - d_{\rm min}^2} \left[ d_{\rm max}^2 \exp\left(i\frac{t}{d_{\rm max}^2}\right) - d_{\rm min}^2 \exp\left(i\frac{t}{d_{\rm min}^2}\right) \right. \notag \\
& \hspace{25mm} \left. + t{\rm H}\left(\frac{t}{d_{\rm max}^{2}}\right) - t{\rm H}\left(\frac{t}{d_{\rm min}^{2}}\right) \right],
\end{align}
and ${\rm H}(\cdot)$ is given by (\ref{eq:H}).
\end{lemma}
\begin{IEEEproof}
See Appendix~\ref{app:charAk}.
\end{IEEEproof}

From $\varphi_{\nsfX_N}(t)$, the ccdf of $\nsfX_N$ can be evaluated as follows \cite{gil-pelaez_1951}:
\begin{align}
\label{eq:gil-pelaez}
\overline{F}_{\nsfX_N}(x) &= \frac{1}{2} + \frac{1}{\pi} \displaystyle\int\limits_0^\infty \frac{{\rm Im}\{\exp(-itx) \varphi_{\nsfX_N}(t) \}}{t}~ {\rm d}t.
\end{align}
\begin{remark}
We have chosen to characterize $\nsfX_N$ by its ccdf instead of its pdf, since the ccdf can be obtained from the characteristic function by evaluating a single integral, which is computationally less intensive than the double integral required to obtain the pdf. Since $\nsfX_N$ is non-negative, the expected value of $h(\nsfX_N)$, for a differentiable real function $h(\cdot)$, can be expressed in terms of $\overline{F}_{\nsfX_N}(\cdot)$, by considering the following relation:
\begin{align}
\label{eq:hexp}
 h(\nsfX_N) &= h(0)+\displaystyle\int\limits_0^{\nsfX_N} h'(u){\rm d}u \notag \\
 &= h(0) + \displaystyle\int\limits_0^{\infty} h'(u)\mathbbm{1}(\nsfX_N>u) {\rm d}u,
\end{align}
where $h'(\cdot)$ denotes the derivative of $h(\cdot)$. Thus, by applying the expectation operator on both sides of (\ref{eq:hexp}), we obtain
\begin{align}
 \nbbE[h(\nsfX_N)] &= h(0)+ \displaystyle\int\limits_0^\infty h'(u) \overline{F}_{\nsfX_N}(u) {\rm d}u.
\end{align}
\end{remark}
While the marginal distribution of $\nsfX_N$ is fairly tractable, as it is the sum of $N$ independent and identically distributed (iid) random variables, the same cannot be said of $\overline{F}_{\nsfY_N|\nsfX_N}(\cdot|x)$. To illustrate this, consider the expression for $\nsfY_N$, given $\nsfX_N=x$:
\begin{align}
 \label{eq:Yn_xn}
 \nsfY_N &= 1- \frac{1}{x^2}\displaystyle\sum\limits_{k=1}^N \nsfA_k^2 - \frac{2}{x^2}\displaystyle\sum\limits_{j=1}^{N-1} \displaystyle\sum\limits_{k=j+1}^N \nsfA_j \nsfA_k (\cos 2\Theta_j \cos 2\Theta_k + \sin 2\nsfTheta_j \sin 2\nsfTheta_k).
 \end{align}
Let $\underline{\nsfA}^{(N)}=[\nsfA_1,~\cdots,~\nsfA_N]^T$. Given $\nsfX_N=x$, it is easily seen from (\ref{eq:XN}) and (\ref{eq:Yn_xn}) that $\underline{\nsfA}^{(N)}$ is a vector of identically distributed, but \emph{not independent} random variables. Hence, in order to characterize $\overline{F}_{\nsfY_N|\nsfX_N}(\cdot|x)$, the conditional joint distribution of $\underline{\nsfA}^{(N)}$, given $\nsfX_N=x$, is required, which is not easy to express in closed-form. From (\ref{eq:YN}), it is clear that the dependence between $\nsfX_N$ and $\nsfY_N$ is induced by the collection of random variables, $\{\nsfB_{k,N}:k=1,\cdots,N\}$. For the sake of tractability, we remove this dependence by assuming $\nsfB_{k,N} \approx m$, for some $m\geq 0$; furthermore, to obtain a random variable whose second-order statistics match that of $\nsfY_N$, we approximate $\nsfY_N$ as follows:
\begin{approximation}
\label{approx:rwalk}
$\nsfY_N \approx \nsfV_N$, where
 \begin{align}
 \label{eq:VN}
 \nsfV_N &:= v \left(1-m^2 \left[\left(\displaystyle\sum\limits_{k=1}^N \cos 2\nsfTheta_k \right)^2 + \left(\displaystyle\sum\limits_{k=1}^N \sin 2\nsfTheta_k \right)^2 \right] \right),
 \end{align}
with the values of $v$ and $m$ being obtained by moment matching with $\nsfY_N$. 
\end{approximation}

\begin{remark}
For the special case when $\underline{\nsfR}^{(N)}=\nsfR\nb1$, the approximation in (\ref{eq:VN}) reduces to an equality (i.e., $\nsfY_N=\nsfV_N=\nsfW_N$), with $v=1$ and $m=1/N$.
\end{remark}
From (\ref{eq:VN}), the mean and variance of $\nsfV_N$ are given by:
\begin{align}
 \label{eq:mean_VN}
 \nbbE[\nsfV_N] &= v(1-m^2N). \\
 \label{eq:var_YN}
 \sigma_{\nsfV_N}^2 &= \nbbE[(\nsfV_N - \nbbE[\nsfV_N])^2] \notag \\
 &= 2{N\choose 2} v^2 m^4,
\end{align}
where (\ref{eq:mean_VN}) follows as a result of $\underline{\nsfTheta}^{(N)}$ being an iid uniform random vector over $[0,2\pi)$. Equating $\nbbE[\nsfV_N]$ and $\sigma_{\nsfV_N}^2$ with the corresponding quantities for $\nsfY_N$, i.e., $\nbbE[\nsfY_N]$ and $\sigma_{\nsfY_N}^2$, we obtain the following expressions for $m$ and $v$:
\begin{align}
 \label{eq:m_exp}
 m &= \frac{1}{\sqrt{v}} \left( \frac{\sigma_{\nsfY_N}^2}{N(N-1)} \right)^{1/4} ,\\
 \label{eq:alpha_exp}
 v &= \nbbE[\nsfY_N] + N \sqrt{\frac{\sigma_{\nsfY_N}^2}{N (N-1)}}.
\end{align}
However, since $\nsfY_N$ is non-negative, a similar requirement on $\nsfV_N$ imposes the following constraint on $m$:
\begin{align}
 \nsfV_N &\geq 0, \notag \\
 \label{eq:rwalk_insight}
\implies ~ m^2 &\max_{\underline{\nsfTheta}^{(N)}} \underbrace{\left[\left(\displaystyle\sum\limits_{k=1}^N \cos 2\nsfTheta_k \right)^2 + \left(\displaystyle\sum\limits_{k=1}^N \sin 2\nsfTheta_k \right)^2 \right]}_\text{Squared-distance of $N$-step random walk} \leq 1.
\end{align}
The term in square parentheses in (\ref{eq:rwalk_insight}) can be interpreted as the squared-distance of an $N$-step two-dimensional random walk with unit step size; thus, it has a maximum value of $N^2$, obtained when all the steps are in the same direction (i.e., $\nsfTheta_k=\nsfTheta$, for all $k$). Therefore,
\begin{align}
\label{eq:m_constraint}
 0 \leq m \leq 1/N.
\end{align}
From (\ref{eq:m_exp}) and (\ref{eq:alpha_exp}), the upper bound on $m$, given by (\ref{eq:m_constraint}), reduces to the following equivalent constraint on the second-order statistics of $\nsfY_N$:
\begin{align}
\label{eq:m_constraint2}
 \frac{(\nbbE[\nsfY_N])^2}{\sigma_{\nsfY_N}^2}- (N^2-N) \geq 0.
\end{align}
\begin{figure}
 \centering
 \includegraphics[scale=0.8]{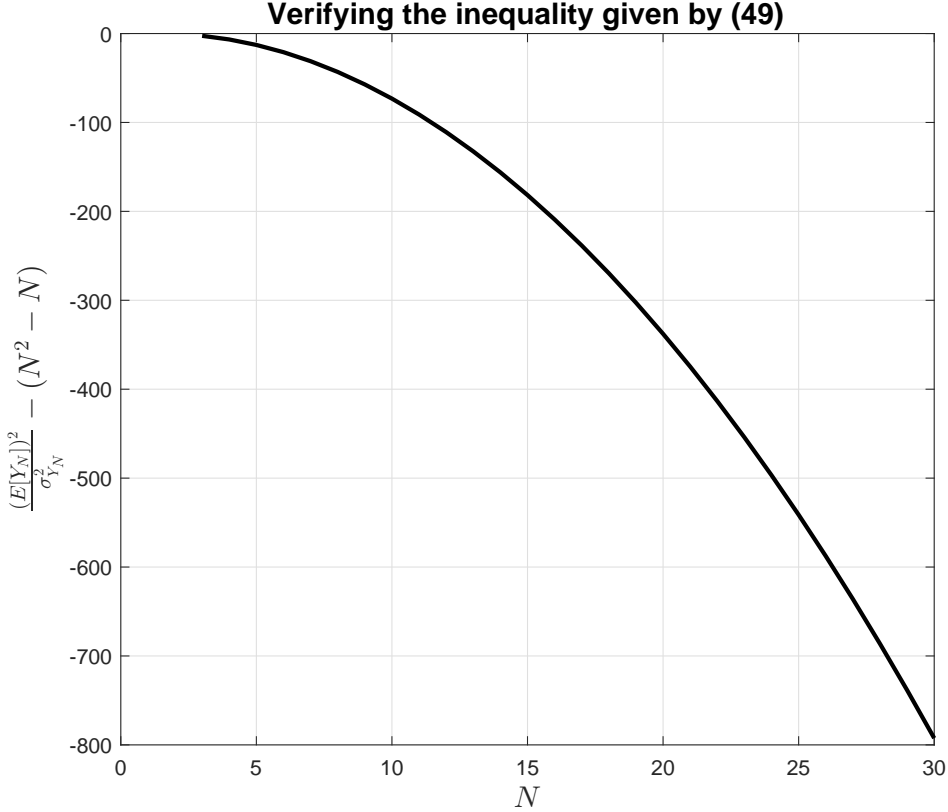}
 \caption{Since (\ref{eq:m_constraint2}) is not satisfied, it follows that (\ref{eq:m_exp}), (\ref{eq:alpha_exp}) and (\ref{eq:m_constraint}) cannot be satisfied simultaneously. The values of $\nbbE[\nsfY_N]$ and $\sigma^2_{\nsfY_N}$ were obtained empirically from $10^6$ samples. For a closed-form characterization of $\nbbE[\nsfY_N]$, see Lemma~\ref{lem:Ymean}.}
 \label{fig:infeasible}
\end{figure}
However, from Fig.~\ref{fig:infeasible}, it can be seen that (\ref{eq:m_constraint2}) is not satisfied for any $N\geq 3$; in fact, the expression on the left-hand side of (\ref{eq:m_constraint2}) becomes increasingly negative as $N$ increases. Thus, it follows that (\ref{eq:m_exp}), (\ref{eq:alpha_exp}) and (\ref{eq:m_constraint}) are not satisfied simultaneously. In particular, the expression for $m$ in (\ref{eq:m_exp}) is greater than $1/N$. As a result, we optimize for the values of $m$ and $v$ in the following manner:
\begin{align}
 \label{eq:obj}
 \min_{m,v} &~ |\sigma^2_{\nsfY_N}-\sigma^2_{\nsfV_N}| \\
 \mbox{subject to}& ~ (\ref{eq:m_constraint}), \notag\\
 \label{eq:mean_match}
 &~ \nbbE[\nsfY_N]=\nbbE[\nsfV_N],
\end{align}
where the above optimization problem can be viewed as \emph{constrained moment matching}, due to the non-negativity constraint on $\nsfV_N$ imposed by (\ref{eq:m_constraint}). From (\ref{eq:mean_match}) and (\ref{eq:mean_VN}), the objective function in (\ref{eq:obj}) can be represented in terms of a single parameter, $m$, as follows:
\begin{align}
 \label{eq:obj_m}
 \bigg|\sigma_{\nsfY_N^2}-N(N-1)(\nbbE[\nsfY_N])^2\frac{m^4}{(1-m^2N)^2} \bigg|
\end{align}
For $m>0$, the expression in (\ref{eq:obj_m}) is initially a monotonically decreasing function of $m$ and attains a minimum value of zero, for $m$ given by (\ref{eq:m_exp}). However, as observed previously, this value of $m$ does not lie in the feasible region, $0\leq m \leq 1/N$. Consequently, the minimum value of (\ref{eq:obj_m}) over the interval $[0,1/N]$ is attained at $m=1/N$. Thus, the optimal solutions for $m$ and $v$, denoted by $m_{\rm opt}$ and $v_{\rm opt}$, respectively, are given by:
\begin{align}
 \label{eq:m_exp2}
 m_{\rm opt} &= 1/N ,\\
 \label{eq:alpha_exp2}
 v_{\rm opt} &:= \frac{\nbbE[\nsfY_N]}{1-m_{\rm opt}^2N},
\end{align}
where the expression for $\nbbE[\nsfY_N]$ is given by the following lemma:
\begin{lemma}
\label{lem:Ymean}
The mean of $\nsfY_N$ is given by
\begin{align}
 \label{eq:EYN}
 \nbbE[\nsfY_N] &= 1-N\left(\frac{\rho_{\rm max}^2-\rho_{\rm min}^2}{2}+\frac{2}{\pi} \displaystyle\int\limits_{\rho_{\rm min}}^{\rho_{\rm max}} \displaystyle\int\limits_{0}^{\infty} u~\frac{{\rm Im}\{\varphi_{\nsfT^{(N)}(u)}(t) \}}{t} ~ {\rm d}t ~{\rm d}u \right) , \\
 \label{eq:splpmin}
\mbox{where}~ \rho_{\rm min} &= \frac{d_{\rm max}^{-2}}{d_{\rm max}^{-2} + (N-1) d_{\rm min}^{-2}}, \\
 \label{eq:splpmax}
 \rho_{\rm max} &= \frac{d_{\rm min}^{-2}}{d_{\rm min}^{-2} + (N-1) d_{\rm max}^{-2}}, \\
\mbox{and}~\varphi_{\nsfT^{(N)}(u)}(t) &= \varphi_{\nsfA_1}((1-u)t)(\varphi_{\nsfA_1}(-ut))^{N-1}.
\end{align}
\end{lemma}
\begin{IEEEproof}
 See Appendix~\ref{app:Ymoments2}.
\end{IEEEproof}

\begin{remark}
Incidentally, note that $\nbbE[\nsfB_{k,N}]=m_{\rm opt}=1/N$. To see this, $\nbbE[\nsfB_{k,N}]$ can be expressed as follows:
 \begin{align}
 \label{eq:mean_BkN}
  \nbbE[\nsfB_{k,N}] &= \nbbE_{\nsfX_N}\left[ \frac{\nbbE[\nsfA_{k}|\nsfX_N=x]}{x}\right], ~ k=1,\cdots,N.
 \end{align}
Since $\underline{\nsfA}^{(N)}$ is a vector of identically distributed random variables, given $\nsfX_N=x$, we have
\begin{align}
 \label{eq:trivial}
 \displaystyle\sum\limits_{k=1}^N \nbbE[A_k|\nsfX_N=x] &= \nbbE[\nsfX_N|\nsfX_N=x] \notag\\
 &=x \notag \\
 &=N\nbbE[\nsfA_k|\nsfX_N=x],~ \mbox{for any k}\in\{1,\cdots,N\} \\
 \label{eq:cond_mean_A1}
 \implies~ \nbbE[\nsfA_k|\nsfX_N=x]&=x/N.
\end{align}
Substituting (\ref{eq:cond_mean_A1}) in (\ref{eq:mean_BkN}), we get
\begin{align}
\label{eq:mean_BkN_final}
 \nbbE[\nsfB_{k,N}]=1/N = m_{\rm opt}.
\end{align}
While, in retrospect, approximating $\nsfB_{k,N}$ by its mean may seem like an obvious choice, the optimality of this approach from a constrained moment matching perspective is not self-evident.
\end{remark}

Substituting $m_{\rm opt}$ and $v_{\rm opt}$ in (\ref{eq:VN}), we obtain
\begin{align}
\label{eq:scaled_WN}
 \nsfV_N &= v_{\rm opt} \nsfW_N.
\end{align}
Using Approximation~\ref{approx:rwalk}, $\SPEB$ can be approximated as follows:
\begin{align}
 \label{eq:speb_newapprox}
 \SPEB &\approx \frac{4}{T_{\rm s} \nsfX_N \nsfV_N},
\end{align}
where $\nsfV_N$ is given by (\ref{eq:scaled_WN}). We now proceed to derive an approximate expression for the ccdf of $\SPEB$ using (\ref{eq:speb_newapprox}).
\begin{theorem}
\label{thm:speb_dist}
The SPEB ccdf, $\overline{F}_{\rm S}(\cdot)$, can be approximated as follows:
\begin{align}
 \label{eq:speb_approxdist}
 \overline{F}_{\rm S}(u) &\approx 1-\nbbE_{\nsfX_N}\left[\overline{F}_{\nsfW_N} \left(\frac{4}{T_{\rm s} u\nsfX_Nv_{\rm opt}}\right) \right] := \overline{F}_{\rm S,app}(u),
\end{align}
where $\overline{F}_{\nsfW_N}(\cdot)$ is given by (\ref{eq:Bessel}), $v_{\rm opt}$ by (\ref{eq:alpha_exp2}) and Lemma~\ref{lem:Ymean}, and the distribution of $\nsfX_N$ by (\ref{eq:gil-pelaez}).
\end{theorem}
\begin{IEEEproof}
From (\ref{eq:speb_newapprox}), we get
\begin{align}
 \label{eq:speb_ccdf_step1}
  \overline{F}_S(u) &\approx 1-\nbbE_{\nsfX_N}\left[\overline{F}_{\nsfV_N|\nsfX_N}\left(\frac{4}{T_{\rm s} ux} \bigg| x\right)\right] \\
  &\overset{(a)}= 1-\nbbE_{\nsfX_N}\left[\overline{F}_{\nsfW_N|\nsfX_N}\left(\frac{4}{T_{\rm s}uxv_{\rm opt}}\bigg| x \right)\right] \\
  &\overset{(b)}= 1-\nbbE_{\nsfX_N}\left[\overline{F}_{\nsfW_N}\left(\frac{4}{T_{\rm s} u\nsfX_Nv_{\rm opt}} \right)\right]\\
  &:= \overline{F}_{\rm S,app}(u),
\end{align}
where $(a)$ follows from (\ref{eq:scaled_WN}), and $(b)$ from the independence of $\nsfX_N$ and $\nsfW_N$. 
\end{IEEEproof}
In the next section, we present numerical results pertaining to Theorem~\ref{thm:speb_dist}.

\section{Numerical Results} \label{sec:NumResults}
For our simulations, we chose $d_{\rm min}=1{\rm m}$ and $d_{\rm max}=10{\rm m}$. For comparison, we consider $\overline{F}_{\rm S}(.)$, obtained from $10^6$ realizations of (\ref{eq:speb_alt}), $\overline{F}_{\rm S,app}(.)$ obtained from Theorem~\ref{thm:speb_dist}, $\overline{F}_{\rm G}(\cdot)$ from (\ref{eq:gdop_tail}) and the ccdfs corresponding to the following \emph{GDOP-based bounds}:
\begin{itemize}
 \item \emph{Upper and lower bounds to $\SPEB$, based on $\GDOP$}: Let $\nsfR_{(1)}$ and $\nsfR_{(N)}$ denote the distance of the nearest and farthest anchors, respectively. $\SPEB$ can then be bounded as follows:
 \begin{align}
  \label{eq:gdop_bounds}
  \nsfG(\nsfR_{(1)},\underline{\nsfTheta}^{(N)}) \leq \SPEB \leq \nsfG(\nsfR_{(N)},\underline{\nsfTheta}^{(N)}).
 \end{align}
As a result, $\overline{F}_{\rm S}(\cdot)$ can be bounded using  (\ref{eq:gdop_tail}), in the following manner:
\begin{align}
 \label{eq:gdop_bds_dist}
1-\nbbE_{\nsfR_{(1)}}\left[ \overline{F}_{\nsfW_N}\left(\frac{4\nsfR_{(1)}^2}{T_{\rm s} Nu}\right) \right] \leq \overline{F}_{\rm S}(u) \leq 1-\nbbE_{\nsfR_{(N)}}\left[ \overline{F}_{\nsfW_N}\left(\frac{4\nsfR_{(N)}^2}{T_{\rm s} Nu}\right) \right], 
\end{align} 
where the ccdfs of $\nsfR_{(1)}$ and $\nsfR_{(N)}$ are given by:
\begin{align}
 \label{eq:r1_ccdf}
 \overline{F}_{\nsfR_{(1)}}(r) &= \left(\frac{d_{\rm max}^2-r^2}{d_{\rm max}^2-d_{\rm min}^2}\right)^N \mathbbm{1}(r \in [d_{\rm min},d_{\rm max}]). \\
 \label{eq:rn_ccdf}
 \overline{F}_{\nsfR_{(N)}}(r) &= \left[1 - \left( \frac{r^2-d_{\rm min}^2}{d_{\rm max}^2 - d_{\rm min}^2}\right)^N \right] \mathbbm{1}(r \in [d_{\rm min},d_{\rm max}]).
\end{align}
\end{itemize}
\begin{figure}
\begin{subfigure}{0.48\textwidth}
 \centering
  \includegraphics[scale=0.5]{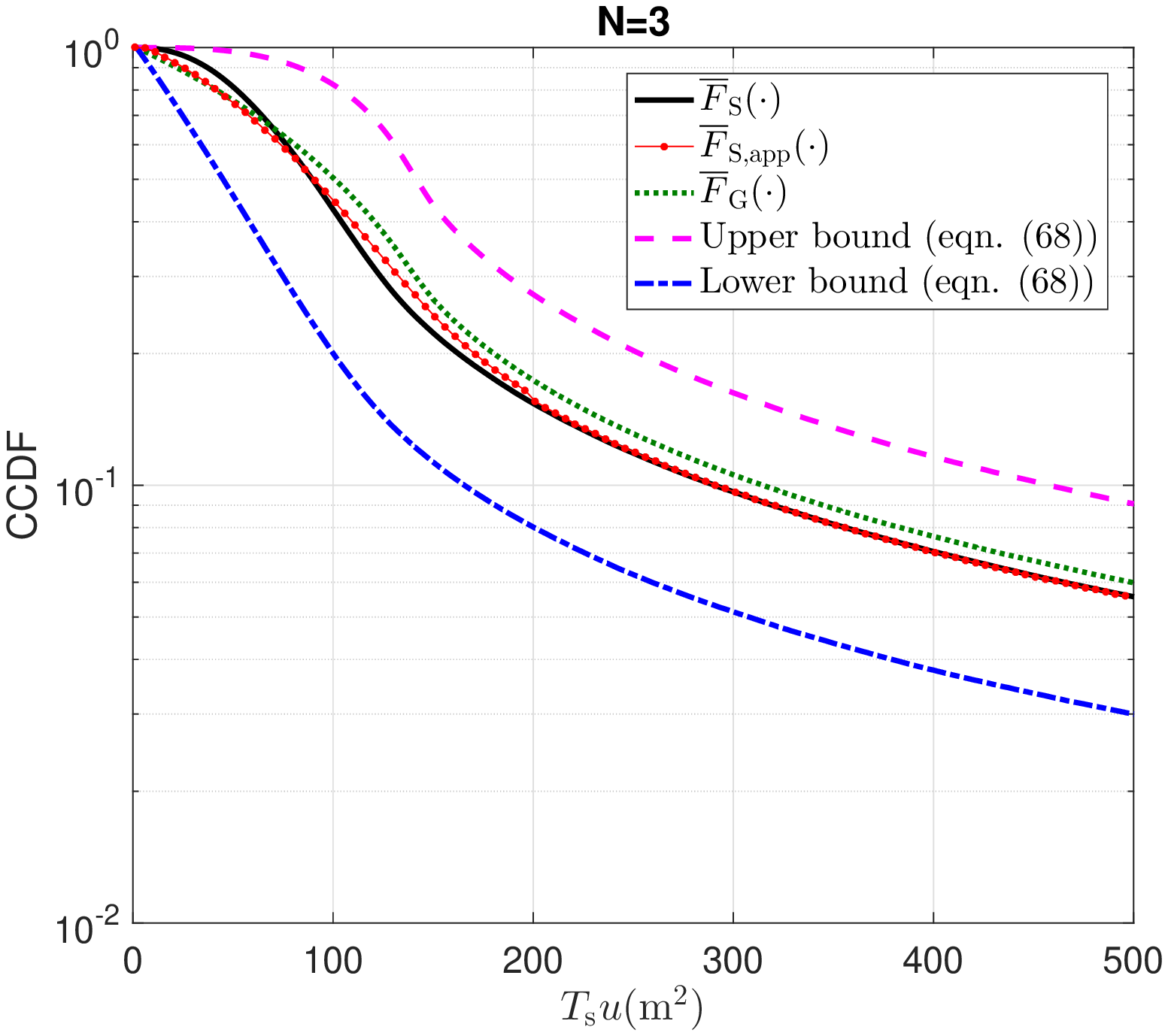}
 \caption{}
 \label{fig:ccdf_N3}
\end{subfigure}
~
\begin{subfigure}{0.48\textwidth}
 \centering
 \includegraphics[scale=0.5]{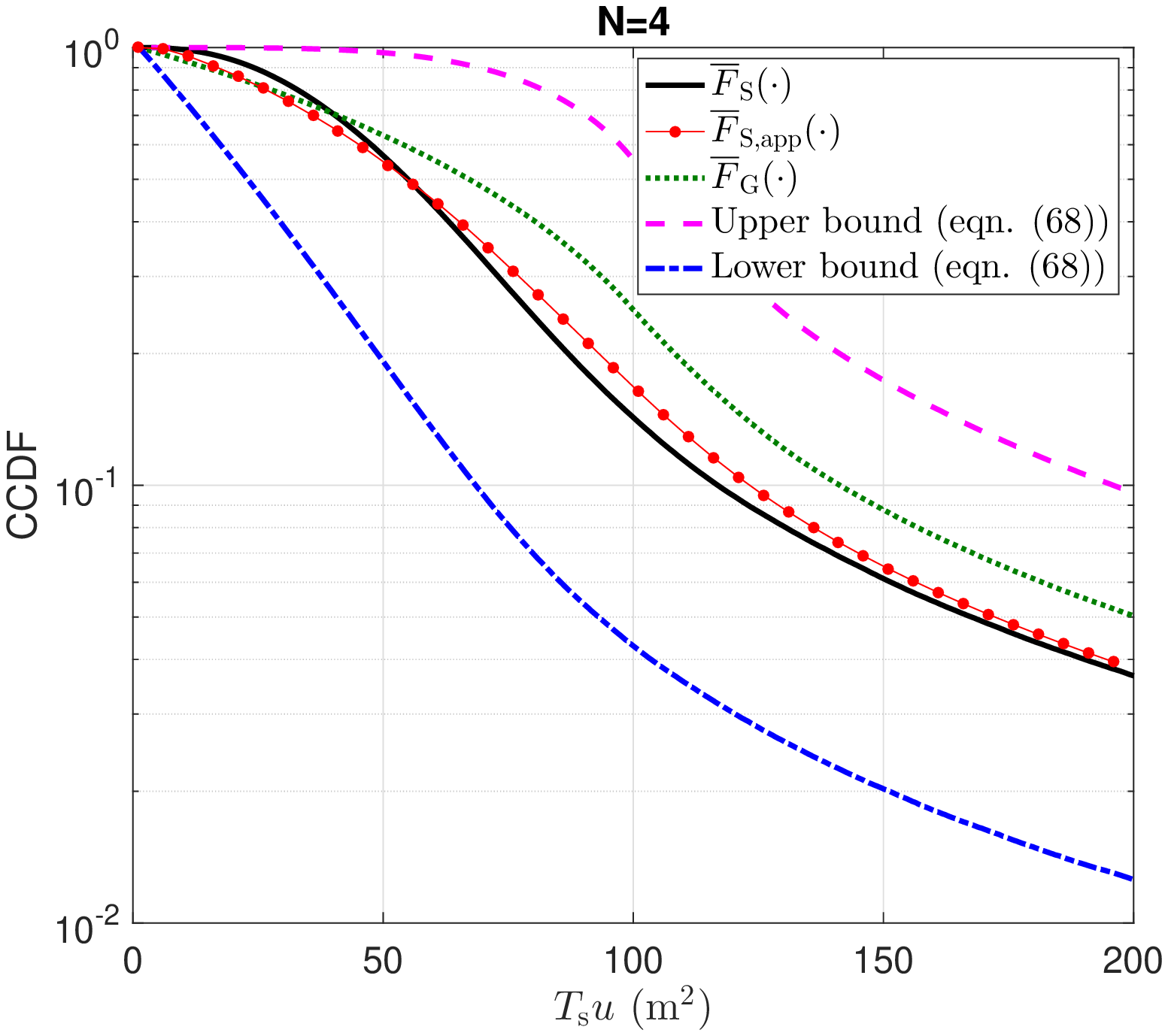}
\caption{}
 \label{fig:ccdf_N4}
\end{subfigure}
\\
\begin{subfigure}{0.48\textwidth}
 \centering
 \includegraphics[scale=0.5]{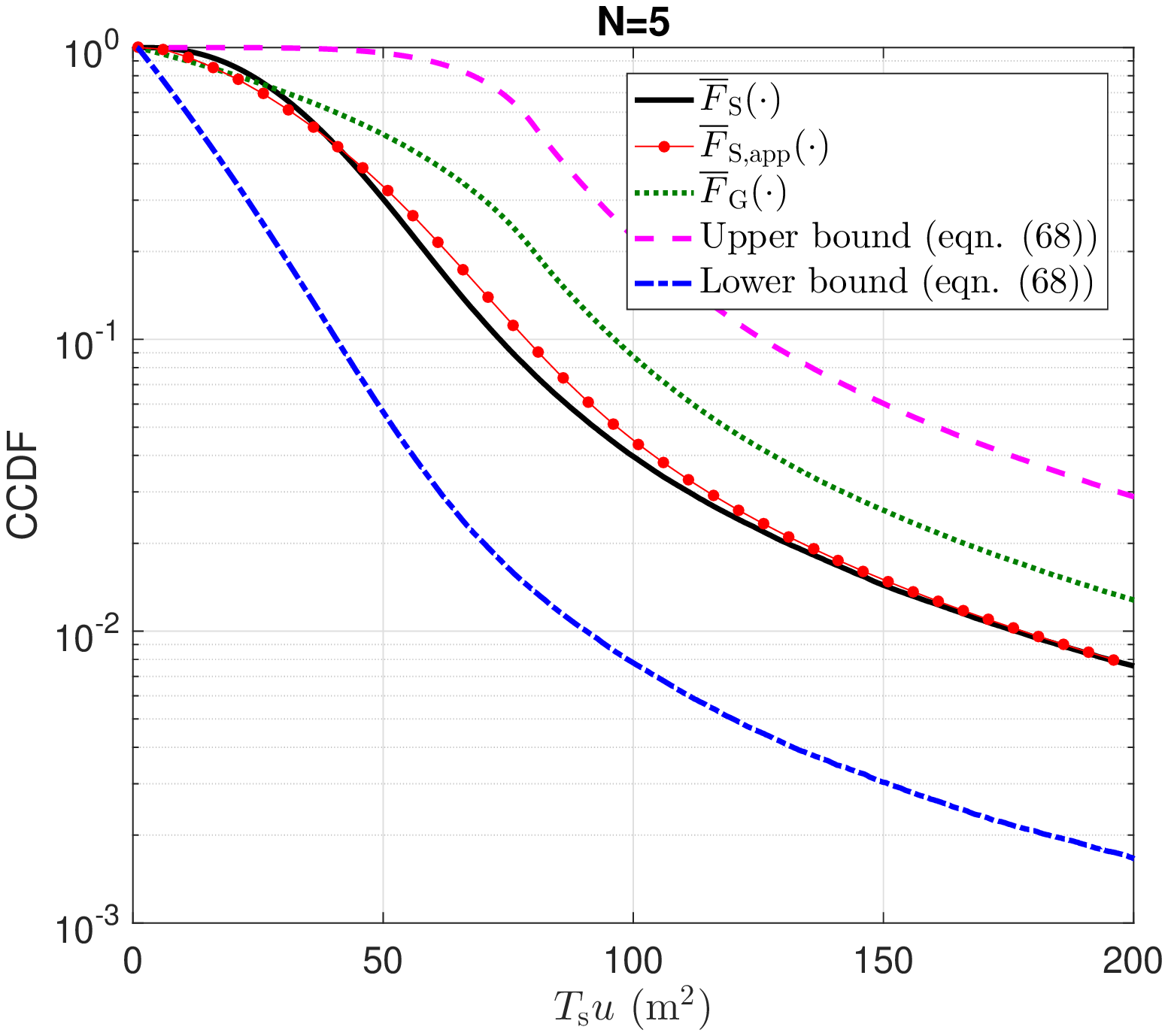}
\caption{}
 \label{fig:ccdf_N5}
\end{subfigure}
~
\begin{subfigure}{0.48\textwidth}
 \centering
 \includegraphics[scale=0.5]{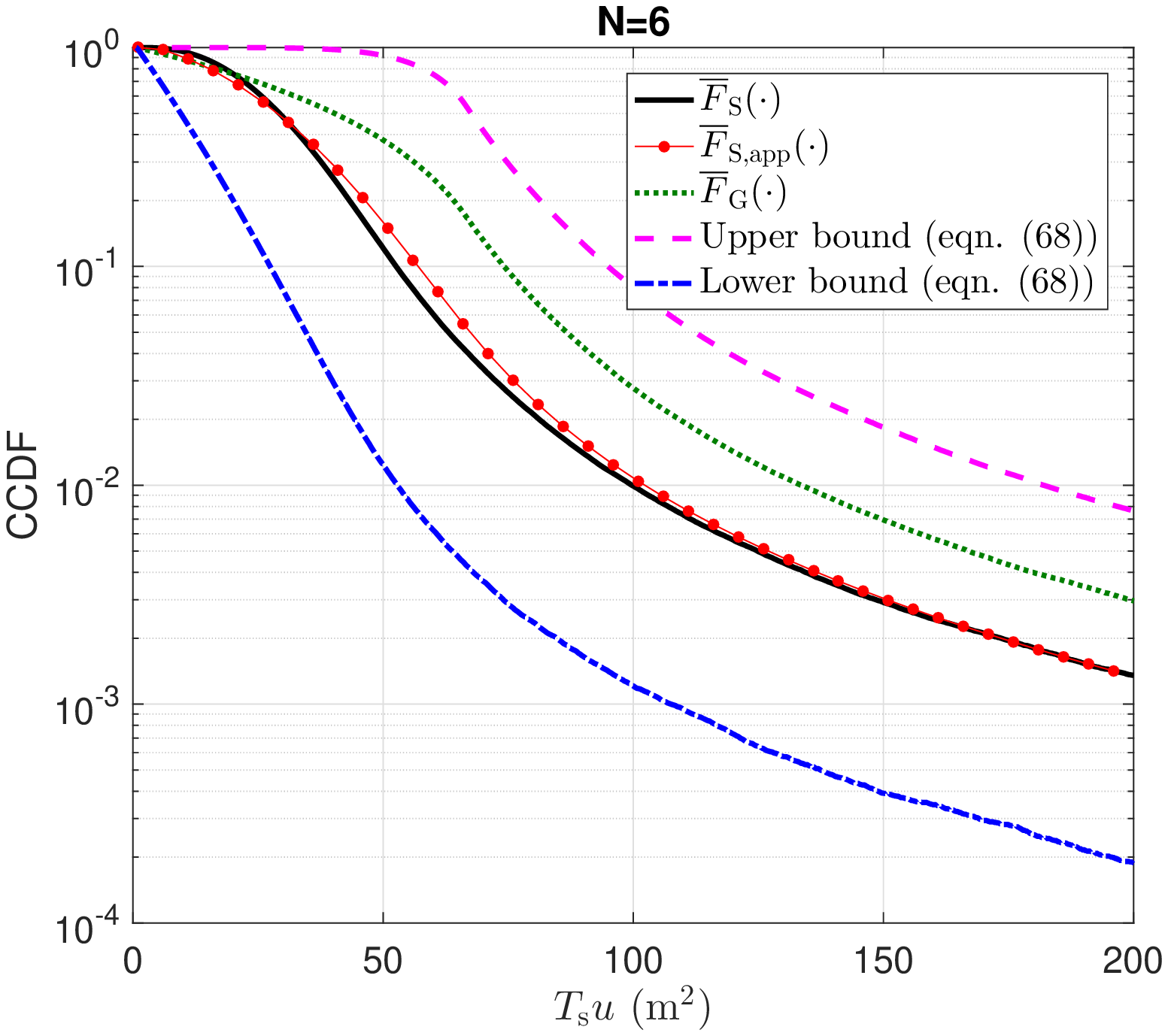}
 \caption{}
 \label{fig:ccdf_N6}
\end{subfigure}
\caption{$\overline{F}_{\rm S,app}(.)$ accurately estimates $\overline{F}_{\rm S}(\cdot)$, which is useful from a design perspective for providing probabilistic guarantees against outage.}
\end{figure}
\begin{figure}\ContinuedFloat
\begin{subfigure}{0.48\textwidth}
 \centering
 \includegraphics[scale=0.5]{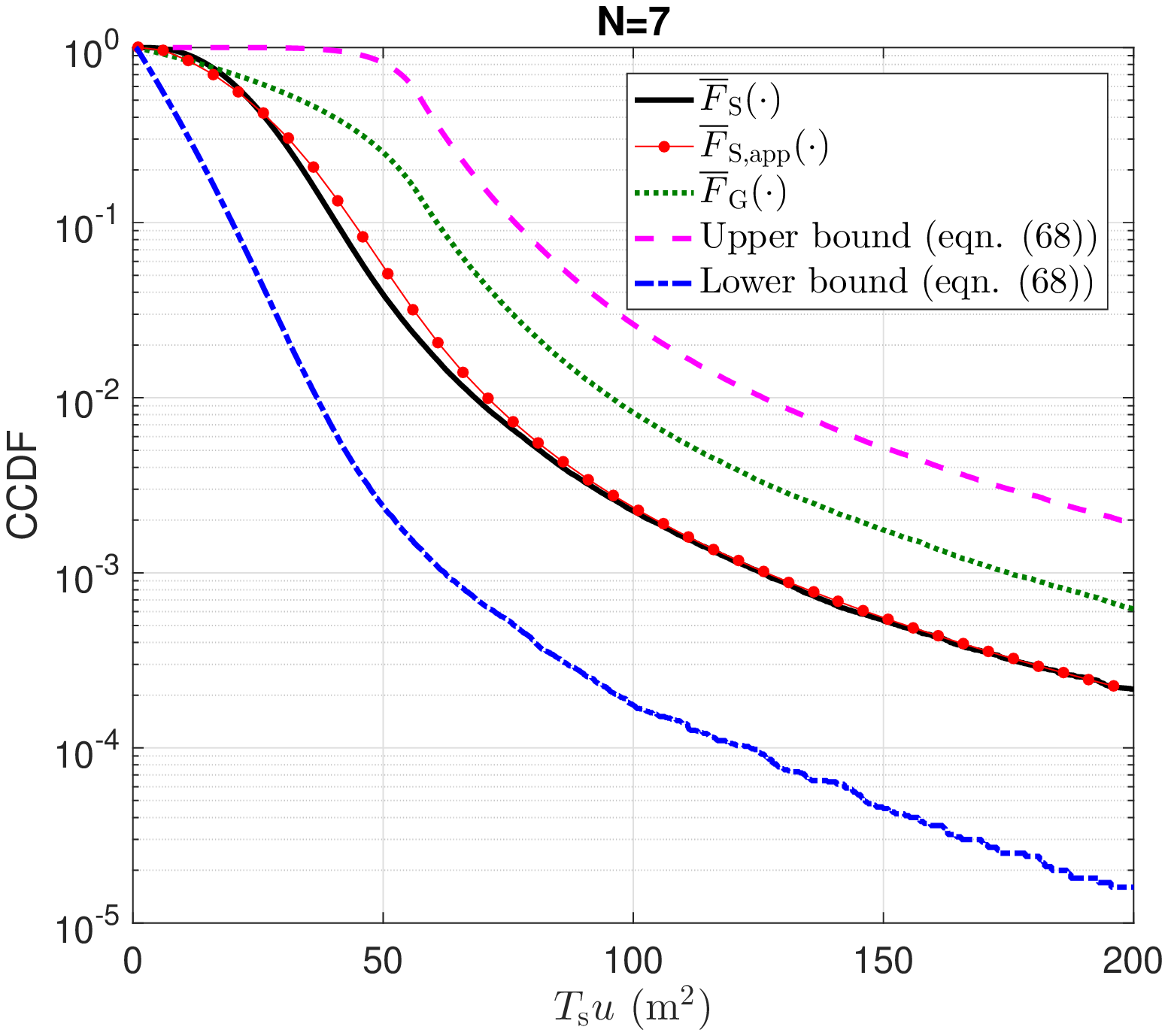}
\caption{}
 \label{fig:ccdf_N7}
\end{subfigure}
~
\begin{subfigure}{0.48\textwidth}
 \centering
 \includegraphics[scale=0.5]{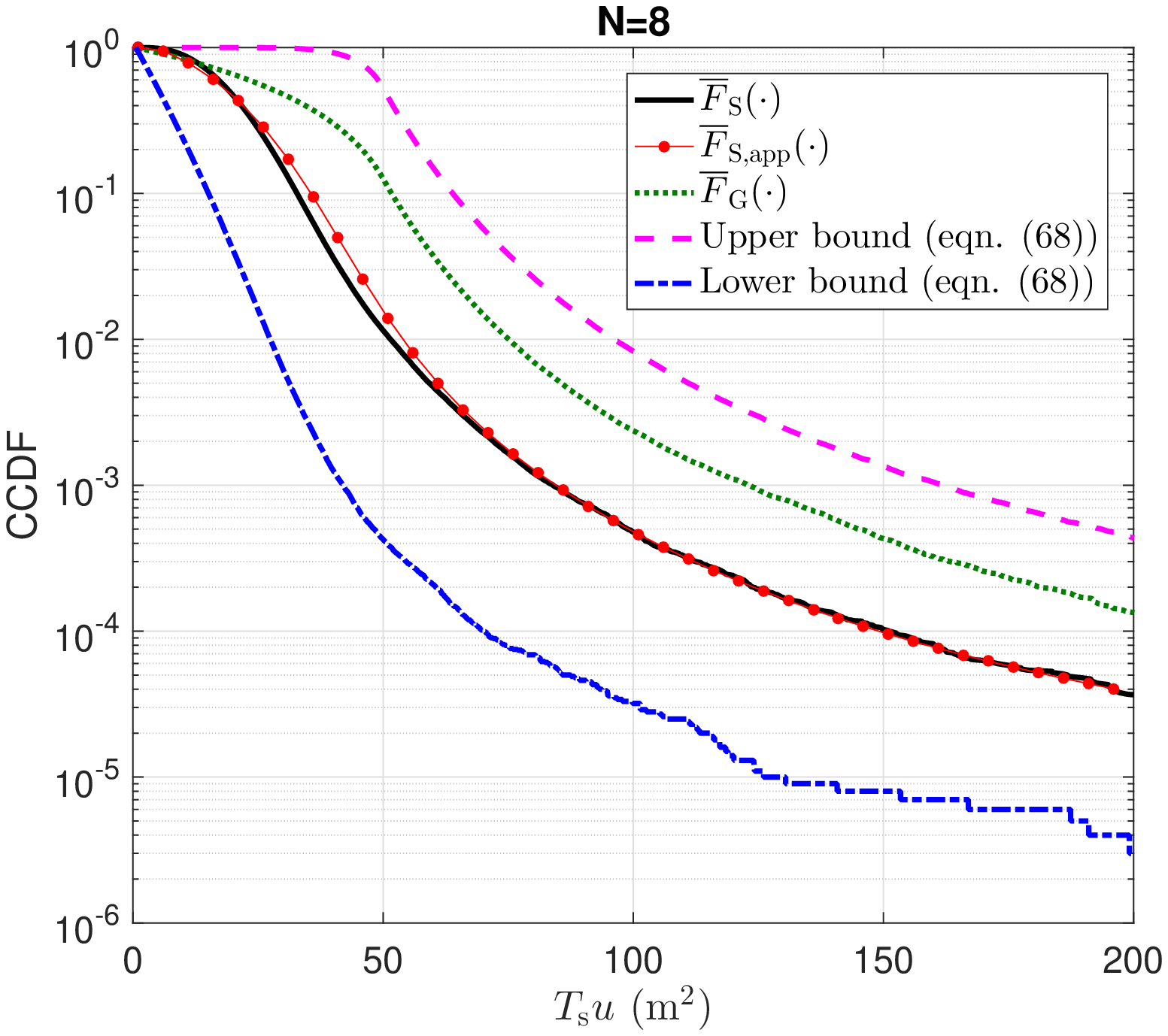}
\caption{}
 \label{fig:ccdf_N8}
\end{subfigure}
 \caption{continued from the previous page.}
 \label{fig:results}
\end{figure}
Like $\GDOP$, $\nsfG(\nsfR_{(1)},\underline{\nsfTheta}^{(N)})$ and $\nsfG(\nsfR_{(N)},\underline{\nsfTheta}^{(N)})$ also have the same support as $\SPEB$. In contrast, for $\nsfH(\underline{\nsfTheta}^{(N)})$, given by (\ref{eq:Chris}), the support mismatch (see Remark~\ref{rem:Chris}) is a significant factor, as the ratio in (\ref{eq:support_mismatch}) evaluates to 21.5 for the values of $d_{\rm min}$ and $d_{\rm max}$ considered. As a result, we do not consider the ccdf of $\nsfH(\underline{\nsfTheta}^{(N)})$ in our analysis.

The ccdf curves are plotted as a function of the SPEB, scaled by the term $T_{\rm s}$, in Fig.~\ref{fig:results} for $N \in \{3,\cdots,8\}$. For all the values of $N$ considered, it can be seen that $\overline{F}_{\rm S,app}$ is accurate at estimating $\overline{F}_{\rm S}(\cdot)$. From a design perspective, the accuracy of $\overline{F}_{\rm S,app}(\cdot)$ at estimating the tail of $\overline{F}_{\rm S}(\cdot)$ is especially useful, as it captures the \emph{outage regime}. Specifically, for outage probabilities below $1\%$, both curves coincide, whereas for a $10\%$ outage probability, the MSE threshold, $\epsilon_{\rm th}$, is slightly larger for $\overline{F}_{\rm S, app}(\cdot)$ than $\overline{F}_{\rm S}(\cdot)$. Consequently, for a given value of $\epsilon_{\rm th}$, $\overline{F}_{\rm S,app}(\cdot)$ can be used to determine the value of $N$ such that the outage probability is at most $10\%$.
\begin{figure}
 \centering
 \includegraphics[scale=0.8]{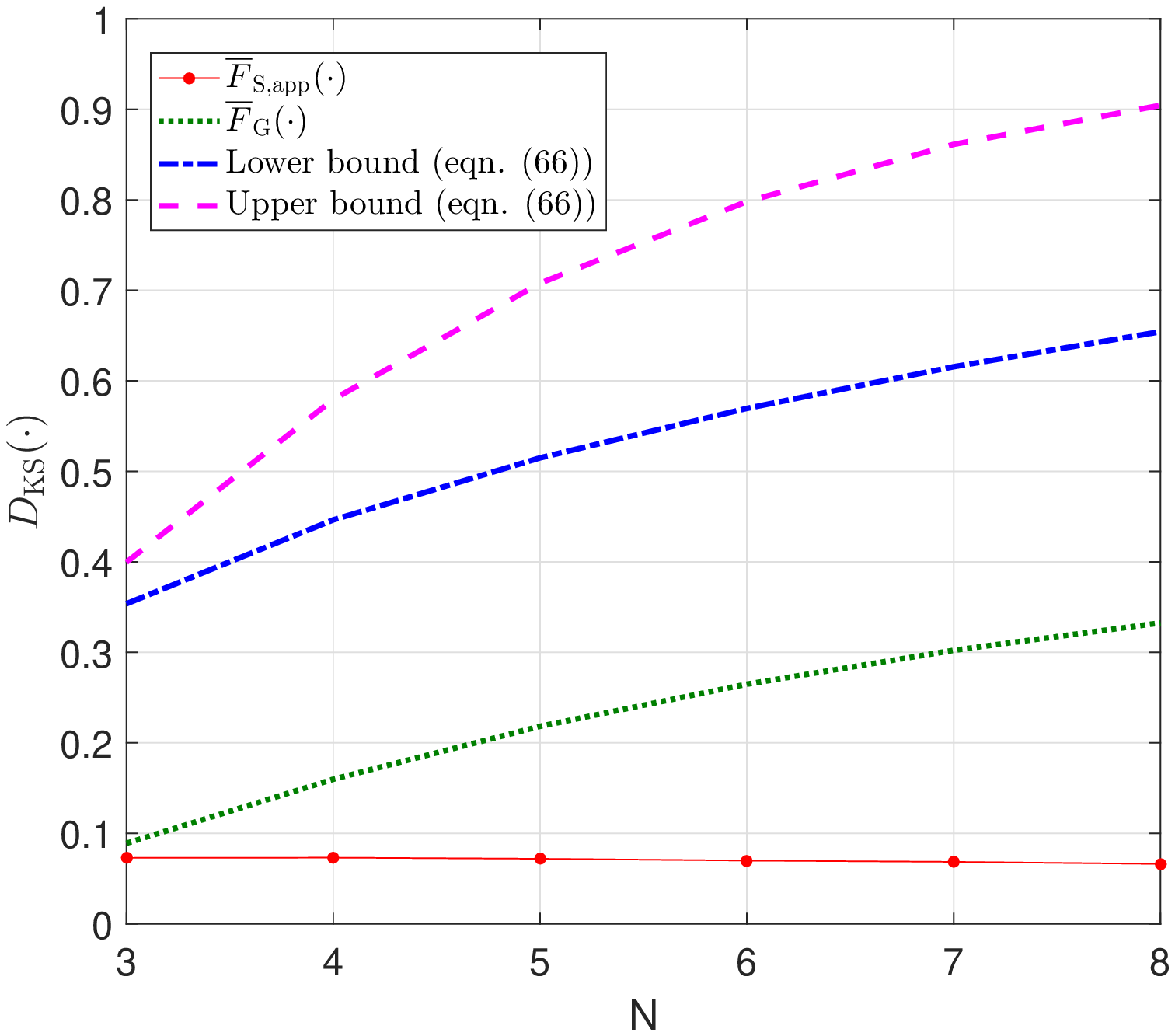}
 \caption{By taking SNR heterogeneity into account, $\overline{F}_{\rm S,app}(\cdot)$ is more accurate at estimating $\overline{F}_{\rm S}(\cdot)$ than the GDOP-based ccdfs.}
 \label{fig:KS}
\end{figure}

\begin{figure}
 \centering
 \includegraphics[scale=0.8]{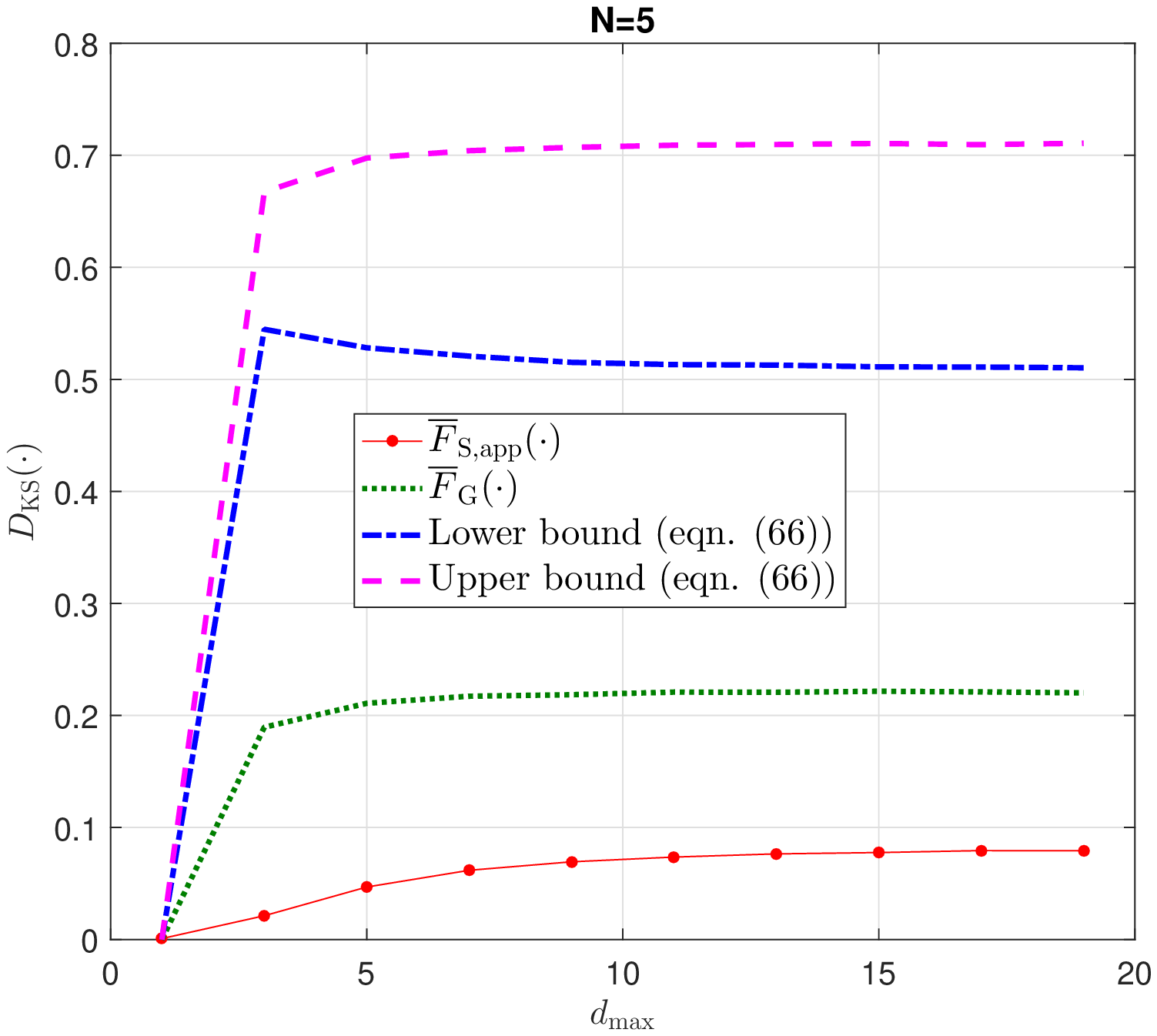}
 \caption{Although $\nsfB_{k,N}\approx \nbbE[\nsfB_{k,N}]=1/N$ becomes less accurate as the difference between $d_{\rm max}$ and $d_{\rm min}$ increases, $\overline{F}_{\rm S,app}(\cdot)$ is still more accurate than the GDOP-based ccdfs, as it takes SNR heterogeneity into account.}
 \label{fig:KS_dist}
\end{figure}
In contrast to $\overline{F}_{\rm S,app}(\cdot)$, we observe that the GDOP-based bounds given by (\ref{eq:gdop_bds_dist}) become progressively loose, while $\overline{F}_{\rm G}(\cdot)$ becomes increasingly inaccurate, as the value of $N$ increases. To quantify this, we use the Kolmogorov-Smirnov (KS) statistic as an error metric, which is defined as follows between a ccdf, $\overline{F}(\cdot)$, and $\overline{F}_{\rm S}(\cdot)$:
\begin{align}
 D_{\rm KS}(\overline{F}) &= \sup_x~ |\overline{F}(x)-\overline{F}_{\rm S}(x)|.
\end{align} 
In Fig.~\ref{fig:KS}, $D_{\rm KS}(\cdot)$ is plotted as a function of $N$, for all the ccdfs considered. Consistent with the insight obtained from Fig.~\ref{fig:results}, we observe that the error increases with $N$ for the GDOP-based ccdfs, while $D_{\rm KS}(\overline{F}_{\rm S,app})$ is nearly constant for all values of $N$. This highlights the importance of considering distance-dependent SNR heterogeneity, especially when the gap between $d_{\rm min}$ and $d_{\rm max}$ is large. To quantify the impact of the difference between $d_{\rm min}$ and $d_{\rm max}$, Fig.~\ref{fig:KS_dist} plots $D_{\rm KS}(\cdot)$ as a function of the $d_{\rm max}$, for $N=5$. While the accuracy of $\overline{F}_{\rm S,app}(\cdot)$ slightly deteriorates with increasing $d_{\rm max}$, as a consequence of $\nsfB_{k,N}\approx \nbbE[\nsfB_{k,N}]=1/N$ becoming less accurate due to the larger variance of $\nsfB_{k,N}$, the resulting error is still smaller than the ones obtained for the GDOP-based ccdfs. Hence, $\overline{F}_{\rm S,app}(\cdot)$ still provides the most accurate estimate of $\overline{F}_{\rm S}(\cdot)$, among previously known approaches.

\section{Summary}
\label{sec:concl}
In this paper, we set out to characterize the impact of distance-based SNR heterogeneity on the error performance of ToA-based localization, using the SPEB metric, $\SPEB$. We considered anchors deployed according to a BPP over an annular region centered around a given target and assumed a distance-dependent inverse-square law pathloss model to capture the SNR heterogeneity. For this setup, $\SPEB$ was shown to be a tightly coupled function of the anchor distances $(\underline{\nsfR}^{(N)})$ and angular positions $(\underline{\nsfTheta}^{(N)})$ and as a result, its ccdf, $\overline{F}_{\rm S}(\cdot)$, was difficult to characterize in closed-form. Hence, we formulated an approximation for $\SPEB$, where the coupling between $\underline{\nsfR}^{(N)}$ and $\underline{\nsfTheta}^{(N)}$ was removed by constrained moment matching, which enabled us to derive a closed-form approximation, $\overline{F}_{\rm S,app}(\cdot)$, of $\overline{F}_{\rm S}(\cdot)$. Through simulations, we observed that  $\overline{F}_{\rm S,app}(\cdot)$ was accurate at estimating $\overline{F}_{\rm S}(\cdot)$, especially at the tail, which corresponds to the outage regime. In particular, from a design perspective, it was observed that $\overline{F}_{\rm S,app}(\cdot)$ can be used to determine the number of anchors needed to guarantee an outage probability of at most $10\%$. Finally, by comparing the accuracy of $\overline{F}_{\rm S,app}(\cdot)$ with GDOP-based ccdfs (obtained by assuming SNR homogeneity) using the KS statistic, $D_{\rm KS}(\cdot)$, we demonstrated that SNR heterogeneity has a considerable impact on $\overline{F}_{\rm S}(\cdot)$.

\appendix

\subsection{Proof of Lemma~\ref{lem:speb_new}}
\label{app:speb_new}

From (\ref{eq:speb_exp}), we have
\begin{align}
 \SPEB &= \frac{\displaystyle\sum\limits_{k=1}^N \nsfR_k^{-2}}{T_{\rm s}\displaystyle\sum\limits_{j=1}^{N-1} \displaystyle\sum\limits_{k=j+1}^{N} \nsfR_j^{-2} \nsfR_k^{-2} \sin^2(\nsfTheta_j - \nsfTheta_k)} \\
 &= \frac{2\displaystyle\sum\limits_{k=1}^N \nsfR_k^{-2}}{T_{\rm s}\displaystyle\sum\limits_{j=1}^{N-1} \displaystyle\sum\limits_{k=j+1}^{N} \nsfR_j^{-2} \nsfR_k^{-2} (1-\cos (2\nsfTheta_j - 2\nsfTheta_k))} \\
 &= \frac{2\displaystyle\sum\limits_{k=1}^N \nsfR_k^{-2}}{T_{\rm s}\displaystyle\sum\limits_{j=1}^{N-1} \displaystyle\sum\limits_{k=j+1}^{N} \nsfR_j^{-2} \nsfR_k^{-2} (1-\cos 2\nsfTheta_j \cos 2\nsfTheta_k - \sin 2\nsfTheta_j \sin 2\nsfTheta_k)} \\
 \label{eq:speb_mod}
 &= \frac{4\displaystyle\sum\limits_{k=1}^N \nsfR_k^{-2}}{T_{\rm s} \left[ \left( \displaystyle\sum\limits_{k=1}^N \nsfR_k^{-2} \right)^2 - \left( \displaystyle\sum\limits_{k=1}^N \nsfR_k^{-2} \cos 2\nsfTheta_k \right)^2 - \left( \displaystyle\sum\limits_{k=1}^N \nsfR_k^{-2} \sin 2\nsfTheta_k \right)^2 \right]},
\end{align}

where (\ref{eq:speb_mod}) is obtained from the following identity
\begin{align}
 \displaystyle\sum\limits_{j=1}^{N-1} \displaystyle\sum\limits_{k=j+1}^{N} a_j a_k &= \frac{1}{2}\left(\displaystyle\sum\limits_{k=1}^N a_k \right)^2 - \frac{1}{2} \displaystyle\sum\limits_{k=1}^N a_k^2 ~, ~a_k\in \nbbR ~ \forall k.
\end{align}
Let
\begin{align}
 \label{eq:Ak_app}
\nsfA_k &= \nsfR_k^{-2}~,~ k\in\{1,\cdots,N\}, \\
 \label{eq:XN_app}
 \nsfX_N &= \displaystyle\sum\limits_{k=1}^N \nsfA_k, \\
 \label{eq:Bkn_app}
 \mbox{and}~ \nsfB_{k,N} &= \frac{\nsfA_k}{\nsfX_N}.
\end{align}
Using (\ref{eq:Ak_app})-(\ref{eq:Bkn_app}), (\ref{eq:speb_mod}) can be expressed as follows:
\begin{align}
 \SPEB &= \frac{4}{T_{\rm s}\nsfX_N \nsfY_N}, \\
\label{eq:YNapp} 
 \mbox{where}~ \nsfY_N &= 1- \left(\displaystyle\sum\limits_{k=1}^N \nsfB_{k,N} \cos 2\nsfTheta_k \right)^2 - \left(\displaystyle\sum\limits_{k=1}^N \nsfB_{k,N} \sin 2\nsfTheta_k \right)^2.
\end{align}
\hfill
\IEEEQED

\subsection{Proof of Lemma~\ref{lem:charAk}}
\label{app:charAk}
Since $\underline{\nsfA}^{(N)}$ is an iid random vector, the characteristic function of $\nsfX_N=\nsfA_1+\cdots+\nsfA_N$ is given by
\begin{align}
\label{eq:charXNapp}
 \varphi_{\nsfX_N}(t) &= (\varphi_{\nsfA_1}(t))^N 
\end{align}
From (\ref{eq:Ak}) and (\ref{eq:fr}), the pdf of $\nsfA_1$ can be expressed as follows:
\begin{align}
\label{eq:pdfAk}
 f_{\nsfA_1}(a) &= (1/2) a^{-3/2}f_{\nsfR}( a^{-1/2}) \\ 
 \label{eq:charAkapp}
\therefore~ \varphi_{\nsfA_1}(t) &= \nbbE[\exp(it \nsfA_1)] \notag \\
 &= \displaystyle\int\limits_{-\infty}^{\infty} \cos(ta)f_{\nsfA_1}(a)~{\rm d}a + i\displaystyle\int\limits_{-\infty}^{\infty} \sin(ta)f_{\nsfA_1}(a)~{\rm d}a \notag\\
 &= I_{\rm c}(t;d_{\rm min},d_{\rm max})+ i I_{\rm s}(t;d_{\rm min},d_{\rm max}), \\
 \label{eq:Ickapp}
 \mbox{where}~ I_{\rm c}(t;d_{\rm min},d_{\rm max}) &:= \displaystyle\int\limits_{-\infty}^{\infty} \cos(ta)f_{\nsfA_1}(a)~{\rm d}a ,\\
 \label{eq:Iskapp}
\mbox{and}~ I_{\rm s}(t;d_{\rm min},d_{\rm max}) &:= \displaystyle\int\limits_{-\infty}^{\infty} \sin (ta) f_{\nsfA_1}(a)~{\rm d}a. 
\end{align}
Integrating (\ref{eq:Ickapp}) and (\ref{eq:Iskapp}) by parts, we get
\begin{align}
\label{eq:Ic2appfinal}
 I_{\rm c}(t;d_{\rm min},d_{\rm max}) &= \left(\frac{1}{d_{\rm max}^2-d_{\rm min}^2}\right) \left[ d_{\rm max}^2 \cos \left(\frac{t}{d_{\rm max}^2}\right) - d_{\rm min}^2 \cos \left(\frac{t}{d_{\rm min}^2}\right) \right. \notag \\
& \hspace{40mm}\left.- t{\rm Si}\left(\frac{t}{d_{\rm min}^2}\right) + t{\rm Si} \left(\frac{t}{d_{\rm max}^2}\right) \right], \\
\label{eq:Is2appfinal}
I_{\rm s}(t;d_{\rm min},d_{\rm max}) &= \left(\frac{1}{d_{\rm max}^2-d_{\rm min}^2}\right) \left[ d_{\rm max}^2 \sin \left(\frac{t}{d_{\rm max}^2}\right) - d_{\rm min}^2 \sin \left(\frac{t}{d_{\rm min}^2}\right) \right. \notag \\
& \hspace{40mm}\left.- t{\rm Ci}\left(\frac{t}{d_{\rm max}^2}\right) + t{\rm Ci} \left(\frac{t}{d_{\rm min}^2}\right) \right],
\end{align}
where ${\rm Si}(\cdot)$ and ${\rm Ci}(\cdot)$ are given by (\ref{eq:Si}) and (\ref{eq:Ci}), respectively.
Combining (\ref{eq:charAkapp}), (\ref{eq:Ic2appfinal}), (\ref{eq:Is2appfinal}) and (\ref{eq:H}), we get
\begin{align}
 \varphi_{\nsfA_1}(t) &= \left(\frac{1}{d_{\rm max}^2-d_{\rm min}^2}\right)\left[d_{\rm max}^2\exp\left(i\frac{t}{d_{\rm max^2}}\right) - d_{\rm min}^2\exp\left(i\frac{t}{d_{\rm min^2}}\right)+ t{\rm H}\left(\frac{t}{d_{\rm max}^2}\right) \right. \notag \\
 & \hspace{40mm} \left.- t{\rm H}\left(\frac{t}{d_{\rm min}^2}\right) \right]. 
\end{align}

\hfill
\IEEEQED

\subsection{Proof of Lemma~\ref{lem:Ymean}}
\label{app:Ymoments2}
$\nsfB_{1,N},\cdots, \nsfB_{N,N}$ form a collection of identically distributed, but \emph{not independent}, random variables. In addition, $\nsfTheta_j$ and $\nsfB_{k,N}$ are also independent random variables, for any $j,k$. Hence, from (\ref{eq:YN}), we obtain
\begin{align}
\label{eq:splEYNapp}
 \nbbE[\nsfY_N] &=  1-N\nbbE[\nsfB_{1,N}^2],
\end{align}
For a fixed $b\in\nbbR$, the function $x/(x+b)$ is increasing in $x$. Hence, from (\ref{eq:Bkn}), it follows that $\nsfB_{1,N}$ is increasing in $\nsfA_1$, for fixed $\nsfA_2,\cdots,\nsfA_N$. As a result, $\nsfB_{1,N} \in [\rho_{\rm min}, \rho_{\rm max}]$, where
\begin{align}
 \label{eq:splpminapp}
 \rho_{\rm min} &= \frac{d_{\rm max}^{-2}}{d_{\rm max}^{-2} + (N-1) d_{\rm min}^{-2}}, \\
 \label{eq:splpmaxapp}
 \rho_{\rm max} &= \frac{d_{\rm min}^{-2}}{d_{\rm min}^{-2} + (N-1) d_{\rm max}^{-2}}. 
\end{align}
As $\nsfB_{1,N}$ is non-negative for all $k$, $\nbbE[\nsfB_{1,N}^2]$ can be expressed as follows: 
\begin{align}
 \label{eq:EB12}
 \nbbE[\nsfB_{1,N}^2] &= 2 \displaystyle\int\limits_{\rho_{\rm min}}^{\rho_{\rm max}} u~ \overline{F}_{\nsfB_{1,N}}(u) {\rm d}u.
\end{align}
From (\ref{eq:Bkn}), $\overline{F}_{\nsfB_{1,N}}(u)$ can be expressed as follows:
\begin{align}
\label{eq:PB1n}
 \overline{F}_{\nsfB_{1,N}}(u) &= \overline{F}_{\nsfT^{(N)}(u)}(0), \\
 \label{eq:T1N}
 \mbox{where} ~ \nsfT^{(N)}(u) &= \nsfA_1(1-u) - u \displaystyle\sum\limits_{j=2}^N \nsfA_j  
\end{align}
Since $\underline{\nsfA^{(N)}}$ is an iid random vector, the characteristic function of $\nsfT^{(N)}(u)$ has the following expression:
\begin{align}
\label{eq:charT1N}
 \varphi_{\nsfT^{(N)}(u)}(t) &= \varphi_{\nsfA_1}((1-u)t)(\varphi_{\nsfA_1}(-ut))^{N-1}.
\end{align}
Similar to (\ref{eq:gil-pelaez}), $\overline{F}_{\nsfT^{(N)}(u)}(0)$ can be evaluated from $\varphi_{\nsfT_1^{(N)}(u)}(t)$, as follows: 
\begin{align}
\label{eq:PTkN}
 \overline{F}_{\nsfT^{(N)}(u)}(0) &= \frac{1}{2} + \frac{1}{\pi} \displaystyle\int\limits_0^\infty \frac{{\rm Im}\{\varphi_{\nsfT^{(N)}(u)}(t) \}}{t} ~ {\rm d}t.
\end{align}
Combining (\ref{eq:splEYNapp})-(\ref{eq:PTkN}), we get
\begin{align}
\nbbE[\nsfY_N]&= 1-N\left(\frac{\rho_{\rm max}^2-\rho_{\rm min}^2}{2}+\frac{2}{\pi} \displaystyle\int\limits_{\rho_{\rm min}}^{\rho_{\rm max}} \displaystyle\int\limits_{0}^{\infty} u~\frac{{\rm Im}\{\varphi_{\nsfT^{(N)}(u)}(t) \}}{t} ~ {\rm d}t ~{\rm d}u \right).
\end{align}
\hfill
\IEEEQED

\bibliographystyle{IEEEtran}
\bibliography{IEEEabrv,JunyangBib,mint_biblio}

\end{document}